\documentclass[traditabstract]{aa}
\graphicspath{./images/}  
\usepackage{lipsum,afterpage}
\usepackage{graphicx}
\usepackage{txfonts}
\usepackage{environ}
\usepackage{booktabs}

\def\cii{[CII]}

\def\mum{$\mu$m}
\def\kms{km s$^{-1}$}
\def\lbol{$L_{\rm Bol}$}

\def\lcii{$L_{\rm [CII]}$}
\def\ergs{erg s$^{-1}$}

\def\lfir{$L_{\rm FIR}$}
\def\msun{M$_\odot$}
\def\msunyr{M$_\odot$ yr$^{-1}$}
\def\lsun{L$_\odot$}
\def\1015{J1015$+$0020}
\def\mjybeam{mJy beam$^{-1}$}
\def\jkms{Jy km s$^{-1}$}

\def\mbh{$M_{\rm BH}$}
\def\mdyn{$M_{\rm dyn}$}
\def\zcii{$z_{\rm[CII]}$}
\def\tacc{$t_{\rm acc}$}
\def\tsfr{$t_{\rm SFR}$}
\def\tedd{$t_{\rm Edd}$}
\def\ttau{$t_{\tau}$}
\begin{document}

   \title{The WISSH quasars project}
   \subtitle{V. ALMA reveals the assembly of a giant galaxy around a z=4.4 \\hyper-luminous QSO \thanks{Based on data from ALMA cycle 4 program 2016.1.00718.S}}

   \author{M. Bischetti \inst{1,2}
                \and E. Piconcelli \inst{1}
                \and C. Feruglio \inst{3}
                \and F. Duras \inst{4}
                \and A. Bongiorno \inst{1}
                \and S. Carniani \inst{5,6}
                \and A. Marconi \inst{7}
                \and C. Pappalardo \inst{8,9}
                \and R. Schneider \inst{10}
                \and A. Travascio \inst{1, 10}
                \and R. Valiante \inst{1}
                \and G. Vietri \inst{11,12}
                \and L. Zappacosta \inst{1}
                \and F. Fiore \inst{3,1}
                }

        \institute{INAF - Osservatorio Astronomico di Roma, Via Frascati 33, I--00078 Monte Porzio Catone (Roma), Italy
                \and Universit\`a degli Studi di Roma "Tor Vergata", Via Orazio Raimondo 18, I--00173 Roma, Italy
                \and INAF - Osservatorio Astronomico di Trieste, via G.B. Tiepolo 11, I--34143 Trieste, Italy
                \and Dipartimento di Matematica e Fisica, Universit\`a degli Studi Roma Tre, via della Vasca Navale 84, I--00146, Roma, Italy
                \and Cavendish Laboratory, University of Cambridge, 19 J. J. Thomson Avenue, Cambridge CB3 0HE, UK
                \and Kavli Institute for Cosmology, University of Cambridge, Madingley Road, Cambridge CB3 0HA, UK
                \and INAF - Osservatorio Astrofisico di Arcetri, Largo E. Fermi 5, I--50125, Firenze, Italy
                \and Centro de Astronomia e Astrof\'isica da Universidade de Lisboa, OAL, Tapada da Ajuda, 1349-018 Lisboa, Portugal
                \and Instituto de Astrof\'isica e Ciencias do Espa\c{c}o, Universidade de Lisboa, OAL, Tapada da Ajuda, 1349-018 Lisboa, Portugal
                \and Universit\`a degli Studi di Roma "La Sapienza", Piazzale Aldo Moro 5, I--00185 Roma, Italy
                \and Excellence Cluster \textit{Universe}, Technische Universit\"{a}t M\"{u}nchen, Boltzmannstr., D-85748 Garching, Germany
                \and European Southern Observatory, Karl-Schwarzschild-Str. 2, 85748 Garching b. M\"{u}nchen, Germany
                }

   \date{28 May 2018}

 
  \abstract
  {We present an ALMA high-resolution ($0.18''\times0.21''$) observation of the 840 \mum\ continuum and [CII] $\lambda157.74$ \mum\ line emission in the WISE-SDSS selected hyper-luminous (WISSH) quasi-stellar object (QSO) \1015, at $z\sim4.4$. Our analysis reveals an exceptional overdensity of [CII]-emitting companions with a very small ($<150$ \kms) velocity shift with respect to the QSO redshift. We report the discovery of the closest companion observed so far in submillimetre observations of high-z QSOs. It is only 2.2 kpc distant and merging with \1015, while two other [CII] emitters are found at 8 and 17 kpc. 
  Two strong continuum emitters are also detected at $<3.5$ arcsec from the QSO. They are likely associated with the same overdense structure of \1015, as they exceed by a factor of 100 the number of expected sources, considering the Log(N)$-$Log(S) at 850 \mum. 
  The host galaxy of \1015\ shows a star formation rate (SFR) of about 100 \msunyr, while the total SFR of the QSO and its companion galaxies is a factor of $\sim10$ higher, indicating that substantial stellar mass assembly at early epochs may have taken place in the QSO satellites.
  For \1015\ we computed a black hole mass $M_{\rm BH}\sim6\times10^9$ \msun. As we resolve the \cii\ emission of the QSO, we can compute a dynamical mass of $M_{\rm dyn}\sim4\times10^{10}$ \msun. This translates into an extreme ratio \mdyn/\mbh$\sim7$, i.e. two orders of magnitude smaller than what is typically observed in local galaxies. The total stellar mass of the QSO host galaxy plus the \cii\ emitters in the ALMA field of view already exceeds $10^{11}$ \msun\ at $z\sim4.4$. These sources will likely merge and develop into a giant galaxy of $\sim1.3\times10^{12}$ \msun. 
  Under the assumption of constant $\dot{M}_{\rm acc}$ or $\lambda_{\rm Edd}$ equal to the observed values, we find that the growth timescale of the host galaxy of \1015\ is comparable or even shorter than that inferred for the SMBH. }

   \keywords{galaxies:~active --  galaxies:~nuclei -- quasars:~emission lines --quasars:~general -- quasars:~supermassive black holes -- techniques:~imaging spectroscopy}

   \maketitle

\section{Introduction}

The most popular models of AGN-galaxy co-evolution \citep{DiMatteo05,Menci08,Hopkins08} include galaxy interactions (both major and minor mergers) and AGN-driven feedback (i.e. the injection of energy and entropy in the interstellar medium (ISM) trough winds and shocks) among the majors processes driving this phenomenon. 
These two processes are highly correlated, for example galaxy interactions may destabilise the gas and make it available for fuelling both star formation (SF) and nuclear accretion, giving rise to the growth of the central super-massive black hole (SMBH) through luminous AGN phases. The AGN fraction increases in IR-luminous, star-forming sources \citep{Nardini&Risaliti11,Rosario13}, and a correlation between the AGN luminosity (\lbol) and the SF luminosity is observed for a wide range of \lbol\ \citep[][]{Netzer09,Lutz10}. In turn, the AGN can power winds \citep{Fiore17} hampering further SF and nuclear gas accretion (as well as AGN-driven winds).

A linked growth of the SMBH and its host galaxy is observationally supported by the well-known local scaling relations between the SMBH mass (\mbh) and the physical properties of the host galaxy bulge \citep[e.g. see][]{Kormendy&Ho13}, such as the dynamical mass (\mdyn) or the velocity dispersion. Theoretically, such relations can be shaped by merger events \citep[see][and references therein]{Alexander12}, triggering at the same time bursts of nuclear and SF activity \citep{Volonteri15,Gabor16,Angles-Alcazar17b}. 
The \mbh-\mdyn\ relation also indicates that the assembly of the giant galaxies can be probed by observing the QSOs with the most massive SMBHs shining at $z>2$. This provides insights into key evolutionary phases not observable in the local universe and allows the investigation of the hotly debated role of mergers and AGN feedback.
This field of research has been revolutionised by ALMA, thanks to its unprecedented sensitivity and broadband coverage because the QSO emission typically outshines the host galaxy at all wavelengths below few tens of \mum. The most powerful observational tool to study the high-z QSOs host galaxies is the [CII] fine structure line at 157.75 \mum. \cii\ is in fact the strongest line from the cool gas ($T<10^4$ K) and, given its low ionisation potential of 11.3 eV, traces both the neutral and ionised medium. It is also a tracer of SF activity \citep{Maiolino05,Carniani13}.
\cite{Trakhtenbrot17} found a wide variety of host galaxy properties of hyper-luminous QSOs at z$\sim$5 in terms of possible SMBH fuelling mechanisms and SF activity, suggesting that galaxy$-$galaxy interactions may not be a necessary condition for either of the two processes.

\cii\ has been investigated even in the most distant $z>6$ QSOs. These objects typically reside in compact hosts where rotating disks are already in place and intense SF activity of tens to thousands of solar masses is ongoing \citep[e.g.][]{Wang13,Wang16,Cicone15,Diaz-Santos16,Venemans16,Venemans17,Willott17,Decarli18}. 
Recent studies of $z\sim6$ QSOs reveal that they are powered by SMBH at the massive end of the black hole mass function \citep{Jiang07,DeRosa11,DeRosa14,Venemans15,Banados16} and that their hosts are among the brightest and most massive galaxies at these redshifts. According to local relations \citep[e.g.][]{Jiang11}, these sources are therefore expected to assemble stellar masses typical of giant galaxies at $z=0$.
In the \mbh-\mdyn\ plane, most of high-z QSOs lie above the local relation, as they are characterised by very low stellar-to-black hole mass ratios as low as $\sim$10. However, the number of $z>4$ sources with available \mbh\ estimates from single epoch relations is very limited to date, while the bulk of the \mbh\ is still derived from the QSO \lbol\ under the assumption of accretion at the Eddington limit. Furthermore, the majority of the high-z QSOs are still unresolved or only marginally resolved, thus affecting size and dynamical mass estimates with large uncertainty. Accordingly, high-resolution studies are of primary importance.

It is therefore crucial to study the SMBH and host galaxy growth at early epochs, i.e. $z\sim2-4$ when both processes are maximised. Accordingly, we have undertaken the WISE-SDSS selected hyper-luminous (WISSH) QSOs project to study the most powerful AGN in the Universe, which happen to shine at these cosmic epochs \citep{Bischetti17}. Similar to the $z\sim6$ QSOs studied so far, WISSH have $L_{\rm Bol}>10^{47}$ \ergs\ and are powered by accretion onto SMBH of \mbh$\sim10^9-10^{10}$ \msun\ at rates close (or even higher than) the Eddington limit. Such huge luminosities at Eddington regimes are likely triggered by galaxy interactions \citep{Menci14,Valiante14,Valiante16} and drive powerful winds that may affect the whole host galaxy. Therefore, these QSOs are ideal targets to shed light on the AGN-galaxy feeding and feedback cycle. To this purpose, we collected information about the AGN power and the multi-transition presence of nuclear and galaxy-scale winds  from multiwavelength spectroscopy and photometry \citep{Bischetti17,Vietri18}. We also built up the far-infrared (FIR-) to-UV spectral energy distributions (SED) of 14 WISSH QSOs with {\it Herschel} photometry \citep{Duras17} to derive the star formation rate (SFR) in their host galaxy. 

From this WISSH-{\it Herschel} subsample, we selected the QSO \1015\ at $z=4.4$ for a pilot ALMA observing programme aimed to characterise the host galaxy and environment properties of hyper-luminous QSOs. The high redshift and low declination guaranteed that this target would fit the ALMA band 7 well and be observed with good sensitivity. Specifically, we present here the results from a high-resolution ($0.18''\times0.21''$) ALMA observation of the 840 \mum\ and [CII] $\lambda157.74$ \mum\ line emission in \1015.
Throughout this paper, we assumed a $\Lambda\mathrm{CDM}$ cosmology with $\rm H_0$ = 67.3 km s$^{-1}$ Mpc$^{-1}$, $\rm \Omega_\Lambda$ = 0.69, and  $\rm \Omega_M$= 0.31 \citep{Planck16}. 
     
\section{ALMA observation and data analysis}

In this work we present the ALMA Cycle 4 observation (project 2016.1.00718.S) of the WISSH QSO SDSS \1015 (celestial coordinates RA 10:15:49.00, Dec +00:20:20.03). The observation was carried out on 5 March 2016 for a 0.6 hours on-source integration time in C36-5 configuration with a maximum projected baseline of 1396 m.
We used the ALMA band 7 receiver and the frequency division mode of the ALMA correlator. This provided us with four spectral windows of 1.875 GHz width, for a total spectral coverage of 7.5 GHz, with a spectral resolution  of 31.25 MHz. 
A first spectral window was centred at $\sim$352 GHz to cover the expected observed frequency of the \cii\ emission line, given the SDSS DR10 redshift $z_{\rm SDSS} = 4.400$ \citep{Paris14}. A second window was put adjacent to it on the lower frequency side, in the case of a blueshift of the rest-frame UV lines with respect to the systemic emission of the source, i.e. the \cii\ emission. The remaining two spectral windows were placed at higher frequencies to account for the continuum emission.

Visibilities were calibrated with the CASA 4.7.0 software \citep{McMullin07} in pipeline mode by applying the default phase, bandpass, and flux calibrators provided. Images were produced using the CASA task clean with natural weights, a 0.03 arcsec pixel size and a 30 \kms\ channel width, and a final beam size of $0.18''\times0.22''$. The ALMA field of view (FOV) of our observation, defined as the region in which the relative sensitivity is higher than 0.5, is a circular area with a radius of $\sim8.5''$.
The continuum map was obtained by averaging over all the spectral windows and excluding 1.2 GHz around the \cii\ emission.
The continuum emission in the spectral region of the \cii\ line was first modelled by fitting the UV plane of the first two spectral windows by a first degree polynomial and then subtracted to the visibilities. 

For all the sources detected with a signal-to-noise ratio higher than five in the ALMA FOV, continuum and \cii\ flux densities were measured by fitting a 2D Gaussian model to the final map in the image plane (see Sect. \ref{sect:closecomp}). Furthermore, from spectral fitting of the \cii\ emission we derived the parameters of the \cii\ line profile (Sect. \ref{sect:spectra}). For each source, we applied a spectral model including one or two Gaussian emission line components, based on the profile of the \cii\ line. \\


\section{Results from the ALMA observation}
 
        \subsection{Continuum and \cii\ emission maps} \label{sect:closecomp}
        \begin{figure*}
                \centering
                \caption{Continuum and \cii\ line emission maps of \1015\ and the other sources detected in the ALMA FOV. \textit{Top panel:} continuum emission maps of the QSO, Cont1, and Cont2 sources. Contours range from 2$\sigma$ to $8\sigma$ in steps of $1\sigma$ (0.04 \mjybeam), while above $8\sigma$ are in steps of 0.16 \mjybeam. \textit{Middle panel:} spatial distribution of all the detected sources, where coordinates are relative to the QSO location. 
                \textit{Bottom panel:} \cii\ emission maps of the QSO and companion sources CompA, CompB, and CompC. Each map is integrated over the velocity range indicated above the panel.
                Contours are shown as in top panel with $\sigma = 0.060, 0.027, 0.070$ Jy beam$^{-1}$ \kms\ in the left, middle, and right panel, respectively. Dashed contours refer to $-2\sigma$. The ALMA beam is shown as a grey ellipse.}
                \vspace{0.1cm}
                \includegraphics[width=0.8\linewidth]{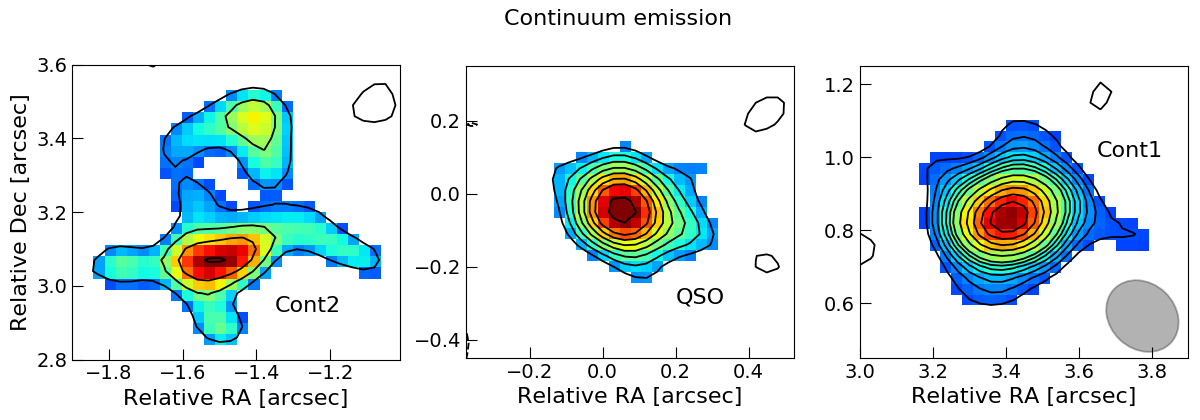}                           \includegraphics[width=0.55\linewidth]{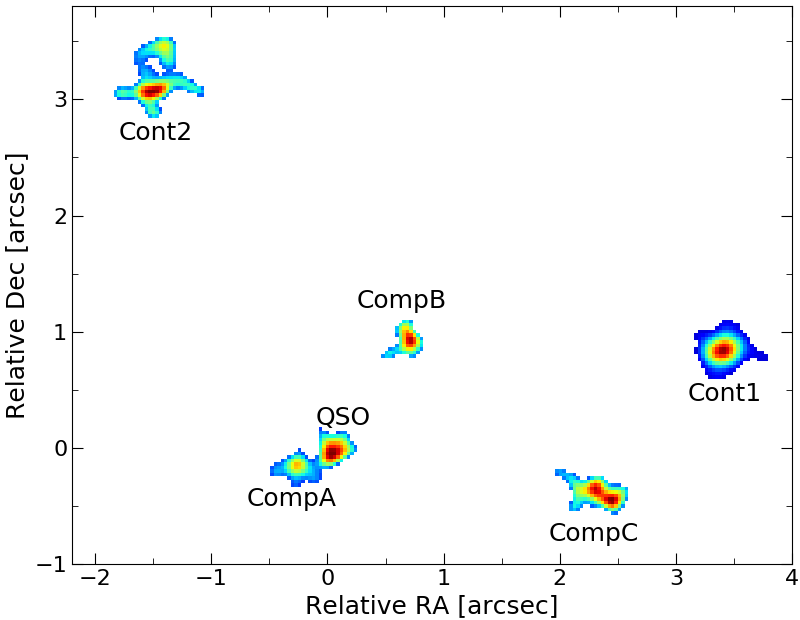}
                \vspace{0.1cm}          
                
                \includegraphics[width=0.8\linewidth]{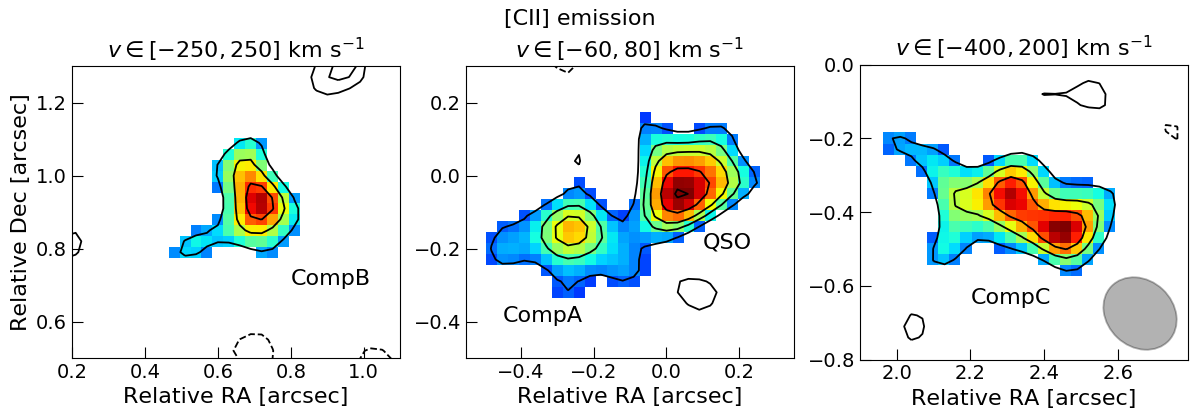}

                \label{fig:cii-map}
        \end{figure*}

        The WISSH quasar \1015\ is detected both in continuum and \cii\ line emission. Furthermore, continuum and \cii\ maps reveal multiple sources with an angular separation of less than 4 arcsec from the central QSO.
        
        \begin{table*}[t]
                \centering
                \caption{Summary of continuum and \cii\ line emission properties of \1015\ and the additional sources detected in the ALMA map. Rows give the following information: (1) source ID, (2) continuum flux, (3)$-$(4) major and minor deconvolved axes of the continuum emission, (5) integrated \cii\ flux density, (6)$-$(7) major and minor deconvolved axes of the \cii\ emission, (8) FWHM of the \cii\ line profile, (9) \cii\ luminosity, and (10) angular separation from \1015. For \cii-only detected sources, i.e. CompA, CompB, and CompC, 3$\sigma$ upper limits on the continuum flux are given. For continuum-only detected sources, i.e. Cont1 and Cont2, we report 3$\sigma$ upper limits on the integrated \cii\ flux density, computed assuming a line width of 300 \kms\ and taking into account the differential sensitivity within the FOV.}
                \begin{tabular}{lcccccc}
                        \toprule
                        (1)\hspace{0.05cm} Source & J1015$+$0020 & Cont1 & Cont2 & CompA & CompB & CompC \\
                        \midrule
                        Continuum emission & & & & & & \\
                        
                        (2)\hspace{0.05cm} $f$ [$\mu$Jy] & 595$\pm$64 & 1252$\pm$76 & 388$\pm$61 & < 120 & < 120 & < 125 \\ 
                        (3)\hspace{0.05cm} $a_{\rm max}$ [$''$] & 0.16$\pm$0.03 & 0.19$\pm$0.02 & 0.28$\pm$0.06 & $-$ & $-$ & $-$\\
                        (4)\hspace{0.05cm} $a_{\rm min}$ [$''$] & 0.94$\pm$0.03 & 0.10$\pm$0.03 & 0.08$\pm$0.06 & $-$ & $-$ & $-$ \\
                        \noalign{\smallskip}\hline\noalign{\smallskip}                  
                        [CII] emission & & & & & & \\
                        (5)\hspace{0.05cm} $f$ [Jy \kms]&       0.47$\pm$0.05 & < 0.15 & < 0.15 & 0.19$\pm0.03$ & 0.35$\pm$0.06 & 1.13$\pm$0.16 \\   
                        (6)\hspace{0.05cm} $a_{\rm maj}$ [$''$] &  0.17$\pm$0.03 & $-$ & $-$ & 0.28$\pm$0.05 & $-$ & 0.59$\pm$0.19 \\
                        (7)\hspace{0.05cm} $a_{\rm min}$ [$''$] &  0.09$\pm$0.03 & $-$ & $-$ & 0.16$\pm$0.04 & $-$ & 0.19$\pm$0.04 \\
                        (8)\hspace{0.05cm} FWHM [\kms] & 339$\pm$38 & $-$ & $-$ & 60$\pm$13 & 480$\pm$45 & 409$\pm$42 \\
                        (9)\hspace{0.05cm} \lcii\ [10$^8$ L$_\odot$] & 2.9$\pm$0.3 & < 0.92 & < 0.92 & 1.2$\pm$0.1 & 2.2$\pm$0.4 & 7.0$\pm$1.0 \\
                        \noalign{\smallskip}\hline\noalign{\smallskip}
                        (10)\hspace{0.05cm} Angular sep [$''$] & $-$ & 3.5 & 3.5 & 0.33 & 1.2 & 2.4 \\
                        \bottomrule
                \end{tabular}
                \label{tab:emission-prop}
        \end{table*}

        Fig. \ref{fig:cii-map} gives an insight into these crowded surroundings. Specifically, the top panels show the ALMA continuum map at $\sim$840 \mum, with a rms sensitivity of 0.04 \mjybeam, revealing cold dust emission from the host galaxy of \1015\ and two additional sources both at a distance of $\sim3.5$ arcsec (Cont1 and Cont2 hereafter). 
        The continuum emission from the QSO is detected at $\sim9\sigma$ significance with a flux density of $f_{\rm cont}^{\rm QSO}\sim600$ $\mu$Jy. The QSO continuum is resolved by the ALMA beam and has a deconvolved size of $0.16\times0.94$ arcsec (see Table \ref{tab:emission-prop}).
        Cont1 and Cont2 are detected at $\sim16\sigma$ and $\sim6\sigma$ significance with a flux density  $\sim 1250$ and  $\sim390$ $\mu$Jy, respectively. These sources are also resolved, see Table \ref{tab:emission-prop}. Cont1 and Cont2 are not detected in \cii\ line emission and, therefore, we are not able to derive a spectral redshift for these sources. 

    \begin{figure*}[]
        \centering
        \caption{ALMA continuum-subtracted spectra of the \cii\ emission line of \1015\ and its companion \cii\ emitters, extracted from an area of 1.5 to 2 beam areas, according to the source. The continuum-subtracted \cii\ flux density is shown as a function of the relative velocity with respect to the QSO redshift \zcii, indicated by the dashed vertical line. The plotted channel width corresponds to 60 \kms. The red curve represents the best fit to the data of one or two Gaussians model. }
        \includegraphics[width=13.7cm]{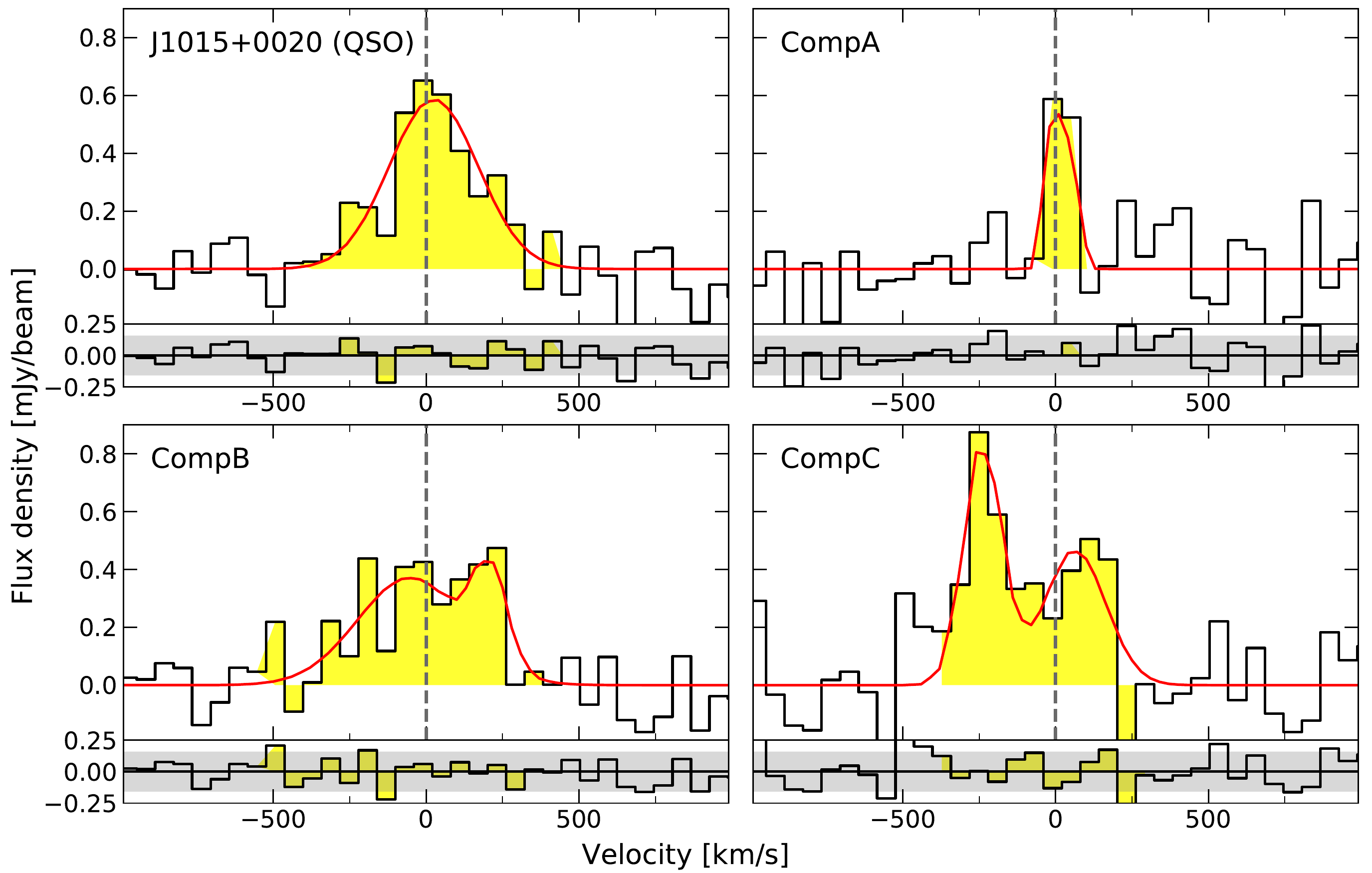}
        
        \label{fig:spectra}
\end{figure*}

        The ALMA \cii\ map also reveals multiple sources around \1015.
    Fig. \ref{fig:cii-map} (bottom middle panel) shows the presence at a $\sim6\sigma$ significance of a clearly distinct companion galaxy with an angular separation of only 0.33$''$ (CompA hereafter), corresponding to a proper distance of $\sim$ 2.2 kpc at the QSO redshift.
    
        Moving to larger distance from the QSO, two additional \cii\ emitters are detected at 1.2$''$(8.2 kpc) and 2.3$''$(16 kpc) with a significance of $6\sigma$ and $7\sigma$, respectively.
        The angular separations from \1015\ are 1.2$''$ and 2.3$''$,  corresponding to $\sim$ 8.2 kpc and 16 kpc, respectively. Hereafter, these sources are referred as CompB and CompC. All three \cii\ emitters are not detected in continuum emission.
        The \cii\ line emission of the QSO is detected at $8.7\sigma$ significance and has an integrated flux density of $f_{\rm [CII]}^{\rm QSO} = 0.47\pm 0.05$ \jkms, derived from 2D-Gaussian fitting.
        CompA, CompB, and CompC have a \cii\ flux density of $\sim$0.19, 0.35, and 1.13 \jkms, respectively.
    The \cii\ emission associated with \1015\ and CompA is marginally resolved by the ALMA beam (see Table \ref{tab:emission-prop}). CompB is unresolved, while CompC has an elongated morphology which clearly rules out a point source nature.

        \subsection{Spectra and velocity maps} \label{sect:spectra}

        We extracted the \cii\ spectrum of \1015, CompA, CompB, and CompC from the ALMA cube in regions of 1.5 to 3 beam areas, according to the size of the source. The resulting continuum-subtracted spectra are shown in Fig. \ref{fig:spectra}. 
        The \cii\ line profile of \1015\ is well reproduced by a single Gaussian component centred at 351.5 GHz, corresponding to a \cii-based redshift $z_{\rm[CII]}=4.407$. This translates into a small velocity shift of about 400 \kms\ between the \cii\ emission and $z_{\rm SDSS}$. 
            \begin{figure*}[t]
                \centering
                \caption{Velocity and velocity dispersion maps of \1015\ (\textit{top panel}) and CompC (\textit{bottom panel}), corresponding to the emitting regions with a signal-to-noise ratio higher than 3 for \1015\ and higher than 2.5 for CompC. Colour bars indicate the velocity and velocity dispersion range of the maps. The ALMA beam is also shown in the right panels as a grey ellipse.  }
                \includegraphics[width=13.7cm]{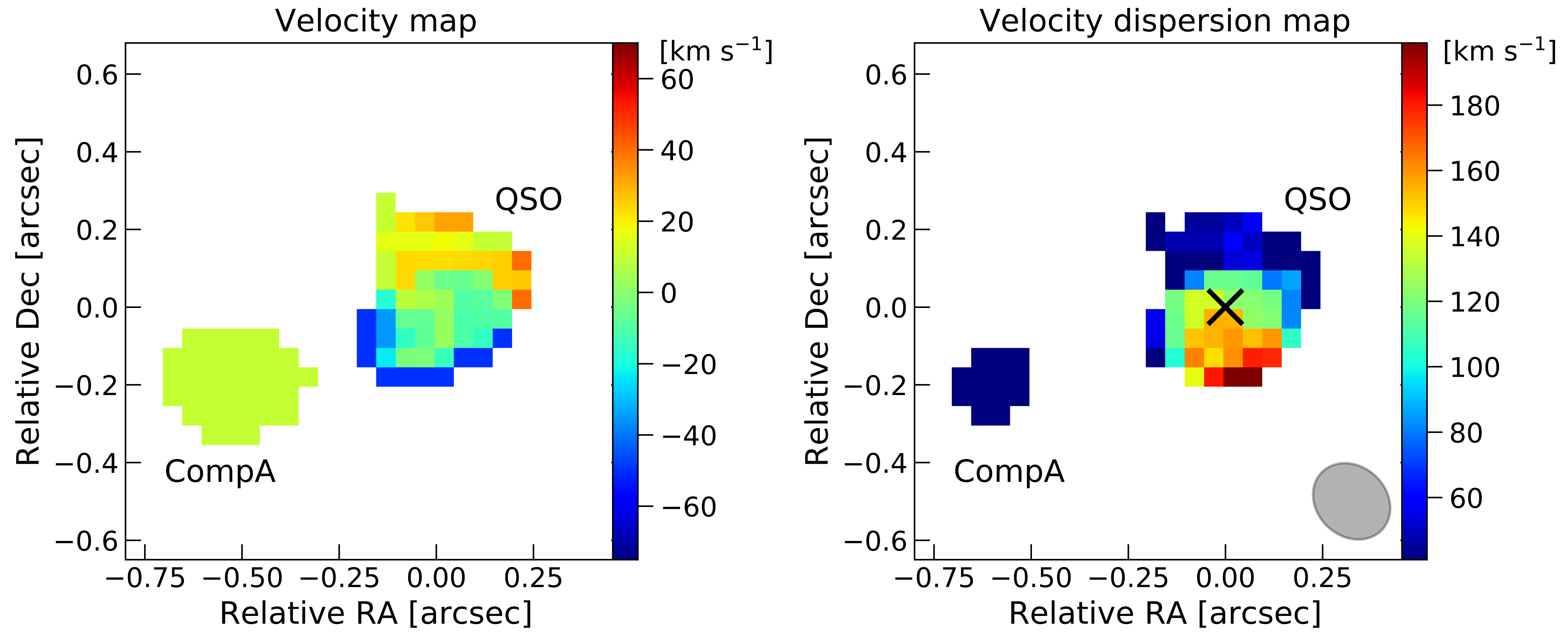}
                \includegraphics[width=13.7cm]{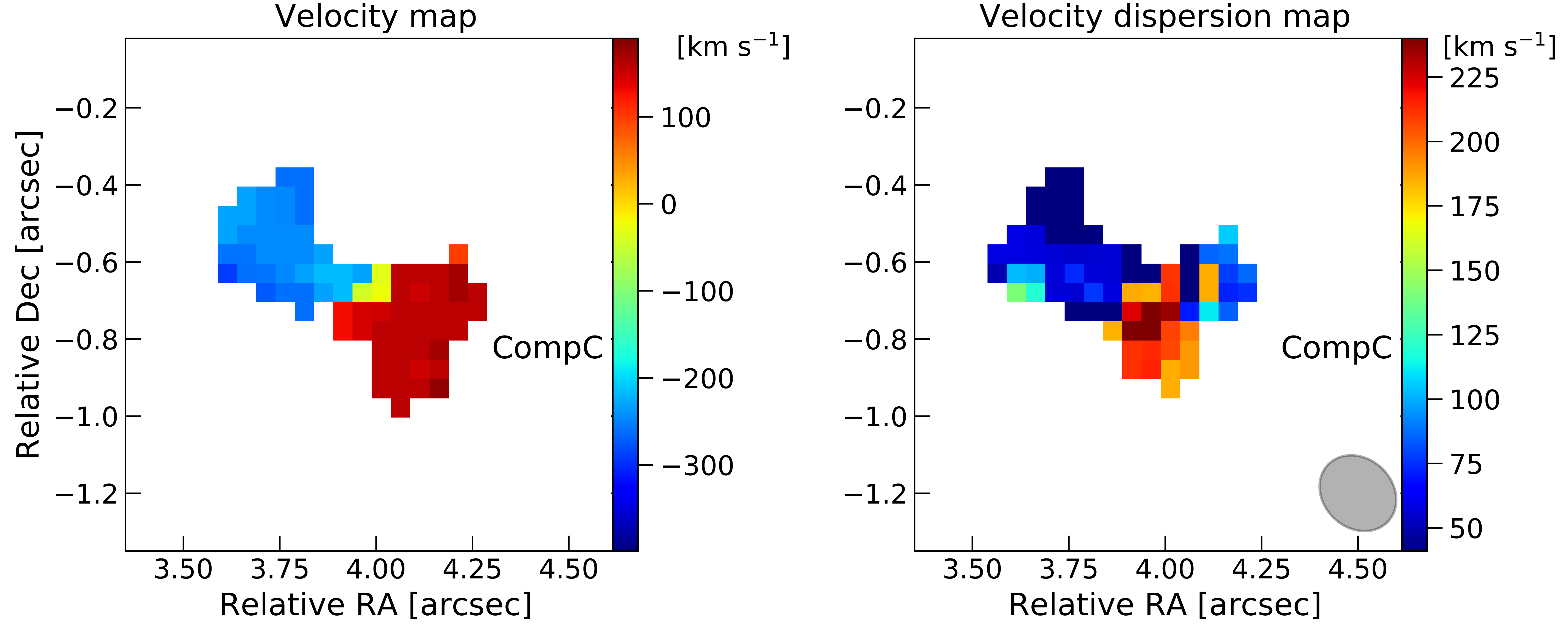}
                
                \label{fig:velmap}
        \end{figure*}

        All the [CII] line-detected sources in the ALMA map have a very small velocity shift of ($\lesssim 150$ \kms) with respect to $z_{\rm[CII]}$ and, therefore, we consider these as companions of \1015. For each source, we calculate the velocity shift as the difference between $z_{\rm[CII]}$ and the velocity that bisects the cumulative \cii\ line flux. CompA, which is the closest companion, has an extremely narrow \cii\ line profile with a FWHM$_{\rm[CII]}^{\rm A} = 60\pm13$ \kms\ centred at the same redshift of the QSO. CompB and CompC are instead characterised by a boxy-double peaked \cii\ line, with  FWHM$_{\rm [CII]}^{B} = 480\pm45$ \kms and FWHM$_{\rm [CII]}^{\rm C} = 409\pm42$ \kms, respectively. These profiles may indicate multiple kinetic components or rotation. The velocity shift of CompB is also consistent with $z_{\rm[CII]}$, while CompC shows a \cii\ line characterised by a blueshift of 145$\pm$30 \kms.

        Fig. \ref{fig:velmap} shows the velocity and velocity dispersion maps of \1015\ (top panel) and CompC (bottom panel). 
        As for the QSO, a small velocity gradient seems to be present in the north-south direction, from about $-60$ to $+40$ \kms, while the \cii\ line of CompA on the QSO left is too narrow to show a gradient at our spectral resolution and appears as a monochromatic spot. 
        The velocity dispersion map of \1015\ clearly shows an increase in the south direction and a peak dispersion of $\sim180$ \kms that is offset with respect to the peak of the \cii\ emission, indicated by the black cross in Fig. \ref{fig:velmap}. An increased central velocity dispersion is usually found at the AGN location \citep[e.g.][]{Trakhtenbrot17}, while we likely observe a perturbation of the gas. This can be interpreted in terms of (i) the very close presence of CompA merging with the QSO, (ii) an additional ongoing, disk scale merger, and (iii) a possible \cii\ outflow component. Deeper observations with higher spatial resolution are needed to draw firm conclusions on the nature of this feature.

        Two blobs with positive and negative velocities with respect to the QSO redshift are present in the velocity map of CompC. The velocity gradient is much larger, ranging from about $-300$ \kms\ to $+150$ \kms.
        We may interpret this as a rotating disk seen at high inclination (see also Sect. \ref{sect:closecomp}) with a diameter of $\sim4$ kpc. Another possibility is that CompC consists of two interacting sources, which are characterised by slightly blueshifted and redshifted \cii\ emission with respect to the $z_{\rm[CII]}$ of the QSO. This is in agreement with the velocity dispersion map, in which the red blob appears more perturbed than the blue blob. However, the quality of the data does not allow us to clearly discriminate between these hypotheses.

\section{Overdensity around J1015$+$0020}\label{sect:over}

\subsection{Multiple companions}
        The ALMA observation has shed light on the crowded surroundings of \1015. We detected three sources, in addition to the QSO, in \cii\ line emission at an angular separation of $\lesssim2.5''$ (see Sect. \ref{sect:closecomp}), corresponding to a proper distance of $\sim17$ kpc at the observed redshift. This high number of sources is very surprising within the context of submillimetre observations of high-z QSOs, where a single companion at larger separation is usually observed \citep{Trakhtenbrot17}.
        Among the \cii\ emitters in our high-resolution observation, we report the discovery of the closest companion observed so far around a high-z QSO. CompA is indeed located at $\sim0.3''$, corresponding to only $\sim2$ kpc from \1015.

        Furthermore, we detected two additional continuum emitting sources separated $\sim3.5''$ from \1015, which have comparable or even higher continuum flux density with respect to the hyper-luminous QSO. Cont1 and Cont2 lack \cii\ line emission in the spectral band covered by our ALMA observation. The non-detection of \cii\ emission from these two sources implies a comoving distance along the line of sight larger than 18.5 Mpc foreground and 25 Mpc background the QSO.
        In order to understand whether they are related to the QSO, we computed the expected number of field sources of any redshift within a region of $4''\times4''$ around the central QSO. By using the Log(N)$-$Log(S) at 850 \mum\ derived by \cite{Simpson2015c}, we should expect $\sim$0.02 field sources with a flux density $\gtrsim0.4$ mJy, i.e. the value observed for Cont2, the faintest continuum source in the ALMA map. The expected number of sources in the same region, with a flux comparable or larger than that of Cont1 is even smaller, i.e. $\sim3\times10^{-3}$. Indeed, assuming a Poissonian distribution with average number of successes equal to the expected number counts, the probability of having two detections is $\sim2\times10^{-4}$. Given the limited sky coverage (about one square degree) of the \cite{Simpson2015c} survey, the observed counts may not be representative of the whole sky. However, as the observed number of continuum sources around the QSO is a factor of 100 larger than the expected value, we conclude that Cont1 and Cont2 are likely associated with the same overdensity traced by the \cii\ emitters around \1015. 
        Recent deep submillimetre surveys carried out with ALMA \citep{Carniani15,Fujimoto16,Aravena16} typically observed $\sim0.1$ sources per ALMA pointing with similar flux to Cont2.
        
        We did not detect any counterpart of the five companions in the Sloan Digital Sky Survey (SDSS), HST/ACS, UKIRT Infrared Deep Sky Survey (UKIDSS), and Wide-field Infrared Survey Explorer (WISE) images in correspondence of their coordinates. 
        \cite{Stark09} and \cite{Bouwens15}  measured typical densities of 0.01 galaxies with SFR$\sim$100 \msunyr\ per ALMA band-7 pointing. Concerning number counts of \cii-emitting galaxies at $z\sim5$, \cite{Aravena16b} predicted about 0.06 galaxies per pointing. 
        In conclusion, it is very unlikely that the presence of three \cii- and two continuum-emitting companions in the field surrounding \1015\ is due to a chance superposition of unrelated galaxies close to the line of sight.
        Accordingly, we are likely observing a very significant overdensity of star-forming galaxies around a powerful and massive QSO.

\subsection{Star formation in and around the QSO host galaxy}\label{sect:SFR}

        \begin{figure*}[t]
                \centering
                \caption{Rest-frame SED of \1015\, before (\textit{left panel}) and after (\textit{right panel}) removing the contamination to the FIR fluxes of the continuum emitters in the ALMA FOV. Black circles indicate the rest-frame photometric points. The black curve represents the total best fit model, while blue and red curves refer to the accretion disk plus torus and cold dust emission, respectively.}

                \includegraphics[width=0.485\textwidth]{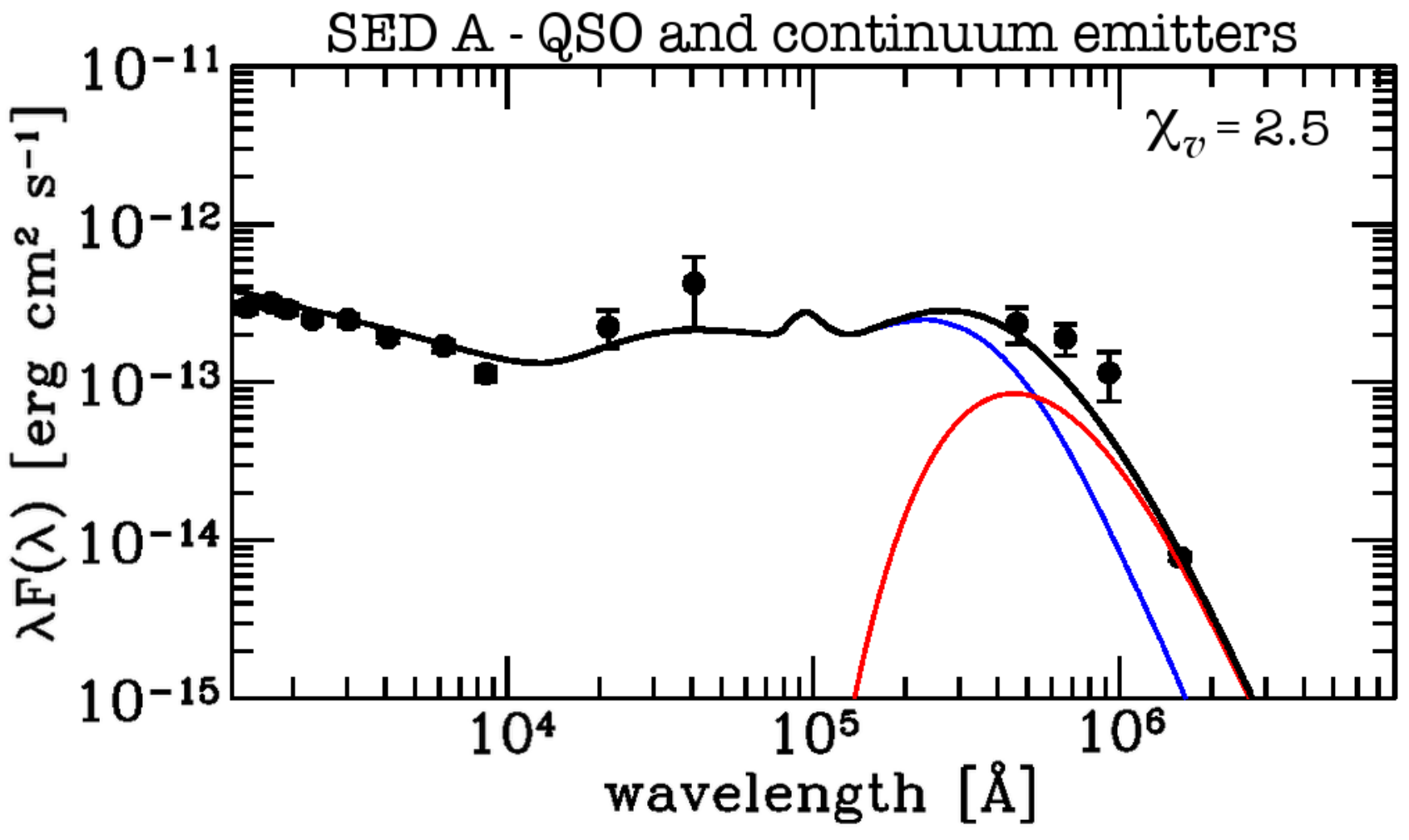}
                \includegraphics[width=0.49\textwidth]{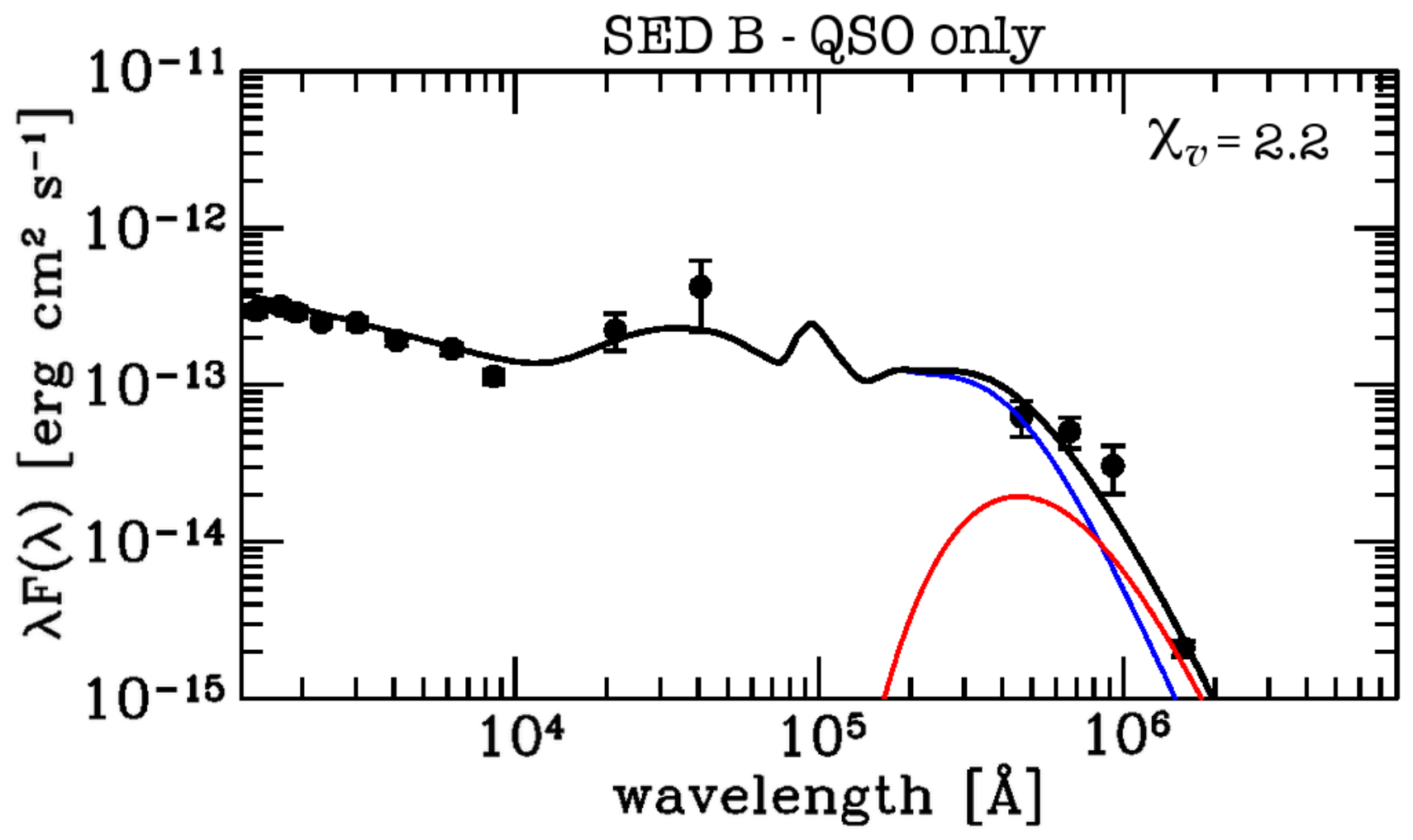}

                \label{fig:sedqso}
        \end{figure*}

        \noindent The ALMA observation allows us to derive the $\sim$ 840 \mum\ continuum flux of \1015\ and, therefore, extend the coverage of the SED presented in \cite{Duras17} to larger wavelengths. 
        Furthermore, ALMA has highlighted the presence of two continuum emitters (Cont1 and Cont2) at an angular separation $\sim 3.5''$ from \1015\ (see Sect. \ref{sect:closecomp}). Both these continuum emitters have a 840 \mum\ continuum flux density comparable to that of the QSO. These sources cannot be resolved as distinct objects in the {\it Herschel} images, given the SPIRE point spread function ranging from $17.6''$ at 250 \mum\ to $35.2''$ at 500 \mum.  
        We therefore performed a new SED fit of \1015\ to remove the contamination of the continuum emitters and more accurately estimate the FIR luminosity (\lfir) of the QSO. We added the ALMA data to the SDSS DR10 \citep{Paris14}, WISE \citep{Wright10}, and {\it Herschel}/SPIRE \citep{Pilbratt10,Griffin10} photometry, presented in \cite{Duras17}. We also included three additional near-IR photometric points from the UKIDSS large area survey \citep{Warren07}.
        In order to quantify the possible contribution of Cont1 and Cont2 to the SPIRE fluxes, we built the following SED:

        \begin{itemize}
        \item SED A, which includes photometric points from SDSS to ALMA 840 \mum. The latter takes into account the continuum emission $f_{\rm cont}^{total}$ from the QSO, Cont1, and Cont2 (Sect. \ref{sect:closecomp}).
        \item SED B, which provides an estimate of the emission from the QSO by removing contamination from Cont1 and Cont2. Specifically, (i) we verified a negligible contribution from the two continuum emitters to the photometric points at wavelength $\le22$ \mum; (ii) we rescaled the SPIRE fluxes by a factor $r_{\rm cont}=3.76$; the latter represents the ratio of $f_{\rm cont}^{total}$ to the ALMA-based flux of the QSO $f_{\rm cont}^{\rm QSO}$ listed in Table \ref{tab:emission-prop}; and (iii) we considered $f_{\rm cont}^{\rm QSO}$ as the flux density for the photometric point at 840 \mum.
        \end{itemize}

        \noindent The rest-frame SEDs A and B are shown in Fig. \ref{fig:sedqso}. Both are well modelled by the sum of an accretion disk plus torus emission component and a cold dust component in the far-IR. However, the SPIRE photometric points show an excess with respect to the best fit model in SED A. This feature is significantly reduced in SED B, indicating that the main contributors to this excess have been removed. Some residual contamination may still be present in particular at 500 \mum because of continuum emitters accounted in the SPIRE fluxes lying outside of the ALMA FOV, i.e. at an angular separation larger than 8.5$''$, which translates to a distance $\gtrsim60$ kpc at the QSO redshift.
        As expected, SED A provides a larger estimate of the FIR luminosity (Log(\lfir/\ergs) = 46.33$\pm0.02$) than SED B (Log(\lfir/\ergs) = 45.69$^{+0.14}_{-0.07}$). Accordingly, the SFR derived by following  \cite{Kennicutt12} is reduced by a factor of 4, i.e. from about 940 to $220^{+68}_{-32}$ \msunyr. 
        The difference with the SFR derived in \cite{Duras17} is even higher, i.e. a factor of 6. This highlights the importance of sampling the FIR and submillimeter bands with high angular resolution in case of high \lbol\ and high-$z$ sources, for which a significantly  enhanced merger rate is expected \citep{Treister12}. 
        Similar results have been indeed reported by \cite{Banerji17} for $z\sim2.5$ heavily-reddened QSOs, pointing out the need for ALMA observations to uncover these structures around luminous QSOs.

        We measured a bolometric luminosity $L_{\rm Bol} = (1.7\pm0.4)\times10^{47}$ \ergs\ (the uncertainty is dominated by that on the QSO inclination).
        We therefore corrected the SFR derived from SED B according to \cite{Duras17}, who found that in hyper-luminous QSOs with Log($L_{\rm Bol}$/erg s$^{-1}$)$ > 47.0$ the AGN contributes to about 50\% of the total FIR luminosity. Accordingly, the resulting SFR of \1015\ is $\sim100$ \msunyr.       
        Throughout the paper, we use quantities derived by SED B as representative of the physical properties of \1015\ and its host galaxy.

        Our ALMA observation reveals that a significant percentage of the 840 \mum\ flux is not associated with the host galaxy of \1015, but instead arises from the surrounding continuum emitters, which likely belong to the QSO overdensity (Sect. \ref{sect:over}). We roughly characterised the spectral shape of Cont1, which is the strongest continuum emitting source in our ALMA map (see Fig. \ref{fig:cii-map}). Cont1 is detected only by ALMA, but SDSS to WISE non-detections can be used to derive upper limits on its flux in all these bands. Specifically, we fitted the star-forming M82 and the starburst Arp220 galaxy templates \citep{Polletta07} to the Cont1 photometry, requiring that they match the ALMA point and leaving the redshift free to vary from $z = 0$ to 5. 
    A M82-like template does not match with the upper limits at any redshift in this range. A $\sim$10 \msunyr\ star-forming galaxy would be in fact visible in the UKIDSS and WISE bands. On the contrary, a more extreme starburst, i.e. an Arp220-like template, is compatible with the SED of Cont1. This suggests that Cont1 is undergoing intense star formation activity of hundreds of \msunyr, in agreement with the FIR excess (corresponding to $\sim700$ \msunyr) observed in SED A associated with the {\it Herschel} photometry. Such a value is derived under the assumption that Cont1 and Cont2 sources belong to the same structure of \1015, which is supported by the very unlikely possibility to have two field sources with such a close angular separation (see Sect. \ref{sect:over}).

    We can estimate the star formation activity of the \cii\ emitters revealed around \1015\ with the relation from \cite{Sargsyan14},    \begin{equation} \label{eq:sfr}
    \mathrm{SFR(M_\odot yr^{-1})} = 1.0\times10^{-7} L_{\rm [CII]}/ \rm L_\odot 
    ,\end{equation}
        which gives SFR within a 50\% uncertainty for low redshift star-forming galaxies. For CompA, CompB, and CompC we derive \cii\ luminosities of $\sim1.2\times10^8$ \lsun, $2.2\times10^8$ \lsun\ , and $7.0\times10^8$ \lsun, respectively (see Table \ref{tab:emission-prop}). According to Eq. \ref{eq:sfr}, these values translate to SFR$^{\rm A}\sim12$ \msunyr, SFR$^{\rm B}\sim22$ \msunyr\,,  and SFR$^{\rm C}\sim70$ \msunyr. We note that, in case of \1015, Eq. \ref{eq:sfr} would correspond to a SFR of only 30 \msunyr, indicating a discrepancy with the \lfir-based value.  This is likely due to the huge radiative power of the AGN in hyper-luminous sources, whose ionising effect reduces the \cii\ emission.

        \begin{figure}[t]
                \centering
                \caption{Ratio of \lcii/\lfir\ as a function of \lfir\ for \1015, compared to high-z QSOs from literature (see Sect. \ref{sect:over} for details) with relative errors. For sources whose uncertainty on \lfir\ was not available, we assume the average value of the sample. \1015\ is indicated by the magenta star, while blue(green) symbols refer to $z\gtrsim6$($z\sim4.5$) QSOs. The best fit relation obtained from orthogonal linear regression and its 1$\sigma$ error are shown by the dashed line and shaded area.}
                
                \includegraphics[width=1\columnwidth]{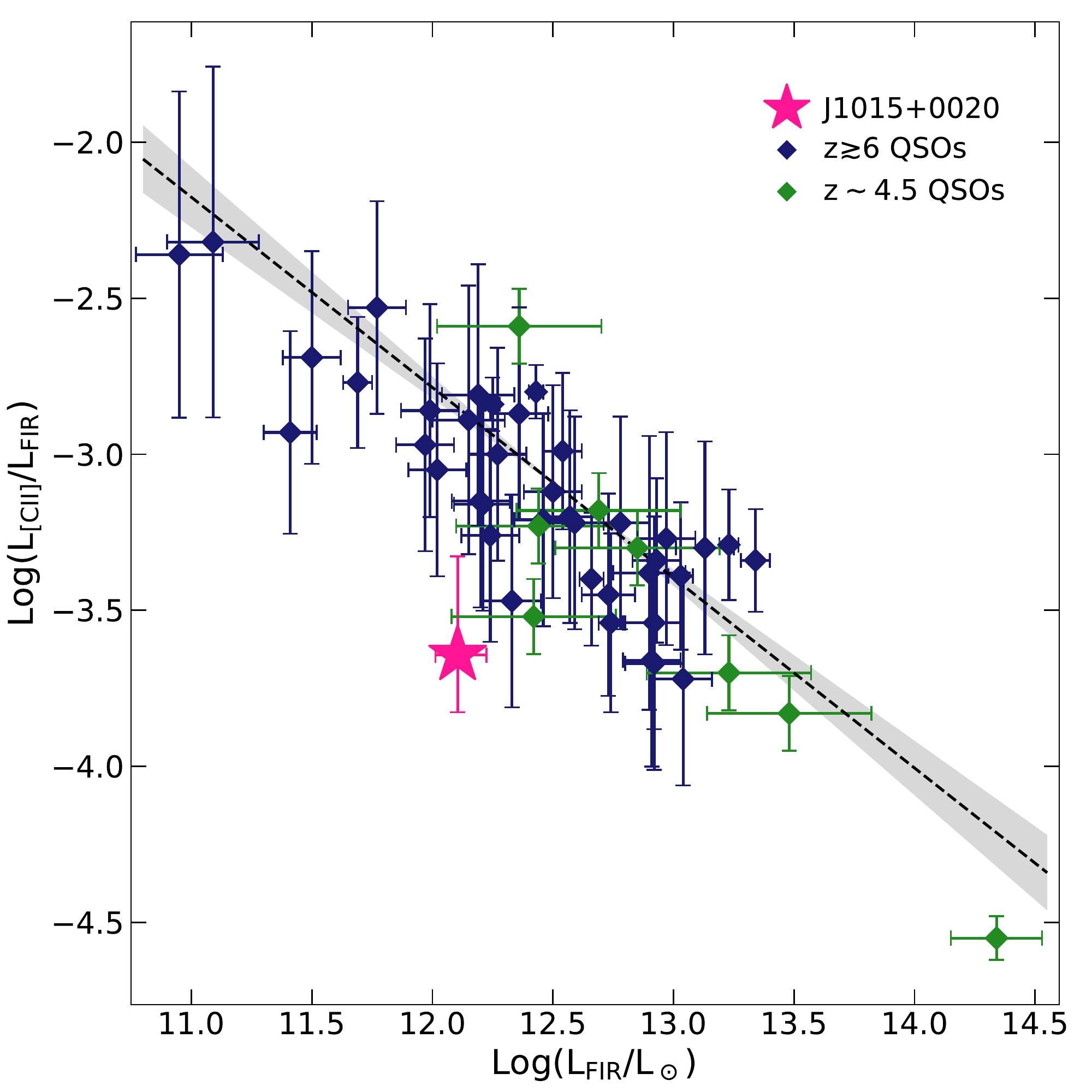}
                \label{fig:lciilfir-lfir}
        \end{figure}
        
        Using the \cii\ flux derived in Sect. \ref{sect:closecomp} and Eq. (1) in \cite{Solomon05}, we compute the \cii\ luminosity of \1015, $L_{\rm [CII]} = (2.9\pm0.3)\times10^8$ \lsun. This implies a Log(\lcii/\lfir) = $-$3.64$^{+0.31}_{-0.18}$, which is among the lowest values found for high-z QSO so far.
        This is shown in Fig. \ref{fig:lciilfir-lfir}, where \1015\ (magenta star) is compared to a sample of $z\sim4-7$ QSOs from literature. Specifically, we collected \lcii\ and \lfir\ for 42 sources from the works of \cite{Wang13,Wang16,Venemans16,Venemans17,Willott13,Willott15,Willott17,Kimball15,Diaz-Santos16,Decarli17}, and \cite{Decarli18}, with relative uncertainties. Whether these are not available, we assume the average uncertainty within the sample. The \lcii/\lfir\ ratio for $z\sim4.5-7$ QSOs span about 1.5 dex and, although with large scatter, a negative trend is evident. We obtain the relation Log(\lcii/\lfir) = $\alpha$Log(\lfir/\lsun) + $\beta$,
        with $\alpha = -0.61\pm0.06$ and  $\beta = 4.5\pm0.8$ by fitting the data with an orthogonal linear regression accounting for errors on both axes. The slope $\alpha$ is slightly steeper than the value of $-0.53$ derived in \cite{Willott17} for $z\sim6$ QSOs. This difference is due to the addition, in our sample, of the hyper-luminous sources from \cite{Trakhtenbrot17} and \cite{Kimball15}, populating the high \lfir\ tail of the sample.

\section{SMBH versus host galaxy properties}\label{sect:smbh-gal}

        \subsection{SMBH and dynamical masses}\label{sect:mdyn}

        We performed a spectral analysis of the rest-frame UV region around the CIV emission line in the SDSS DR10 \citep{Paris14} spectrum of \1015 with the aim of estimating the SMBH mass based on the width of the CIV line profile and luminosity at 1350 \AA.
        
        Black hole masses based on CIV can be affected by a factor of a few to ten uncertainties \citep{Shen&Liu12}. However, in case of \1015\ the profile of this line indicates a small deviation from a symmetric profile associated with Keplerian velocity.
        The velocity shift $\Delta v_{\rm CIV}$ of the CIV line with respect to the [CII]-based redshift of the QSO is indeed moderate ($\sim1000$ \kms) for this AGN luminosity regime, once compared to other WISSH QSOs. \cite{Vietri18} measured the CIV velocity shift with respect to the systemic H$\beta$ emission for a sample of 18 WISSH QSOs, finding values up to 8000 \kms with an average $\Delta v_{\rm CIV}\sim3000$ \kms, which is in agreement with the large shifts typically observed in high-luminosity QSOs \citep[e.g.][]{Sulentic17,Hamann17}. Therefore, we are likely observing the CIV emitting region at large inclination angle, as also indicated by the ALMA observation (see below in this section). According to \cite{Vietri18}, higher line-of-sight inclinations correspond to smaller distortions of the CIV line profile, while low inclinations are associated with very broad, asymmetric profiles due to outflowing gas.
        
\begin{figure}[t]
                \centering
                \caption{Rest-frame UV spectrum of \1015, corresponding to the CIV spectral region. The total best fit model is  shown in red, while the green(blue) curve refers to the CIV core(wing) emission. Continuum emission is shown in purple. The vertical dashed line corresponds to the \zcii\ of the QSO and the grey shaded region indicates the spectral region excluded from the fit because of telluric features.}
                
                \includegraphics[width=1\columnwidth]{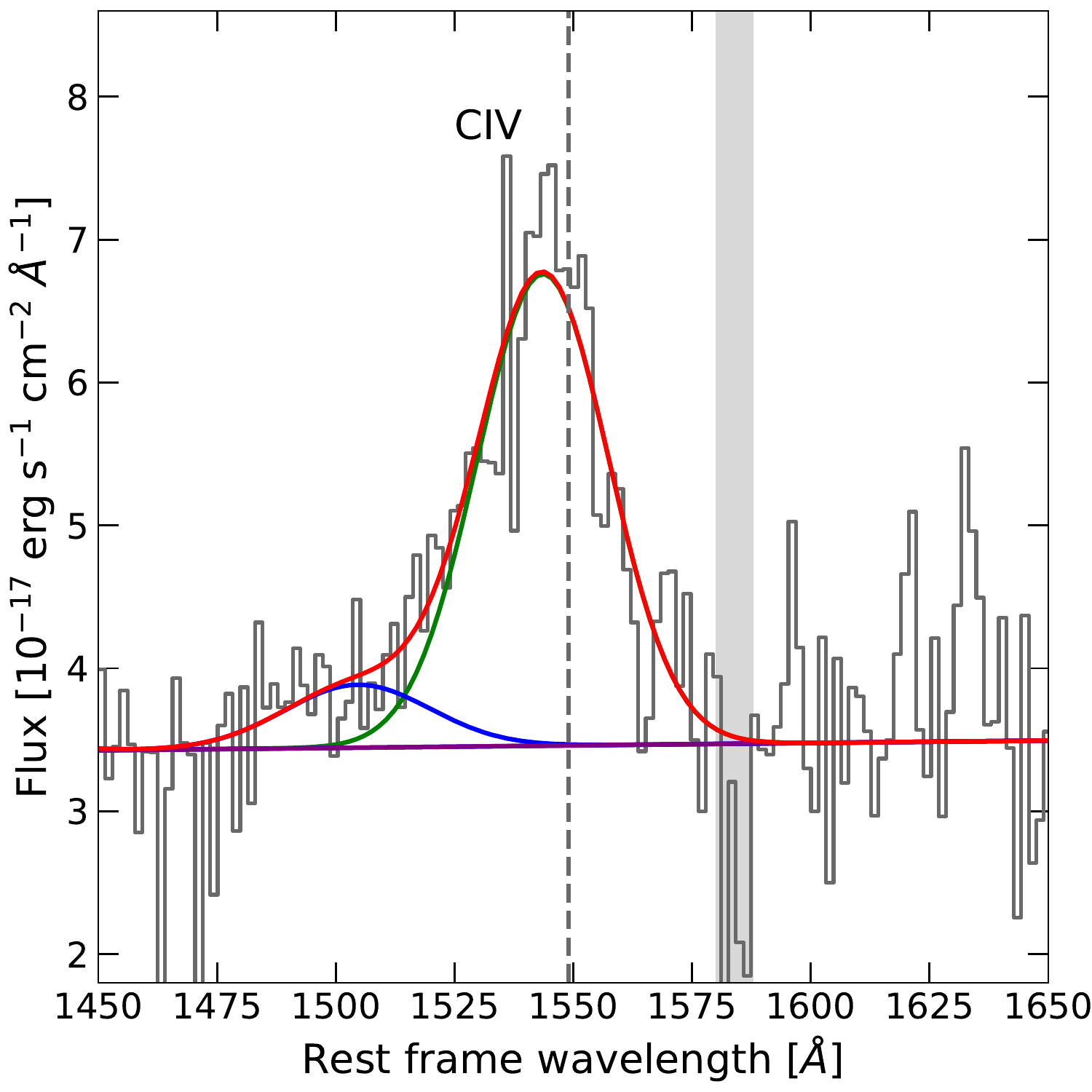}
                \label{fig:civspec}
        \end{figure}

        Specifically, we fit the spectral region between 1300 \AA\ and 1700 \AA\ with a model consisting of a power law to reproduce the continuum emission: one Gaussian component to account for the BLR emission of CIV and a second Gaussian component to fit possible CIV wings associated with outflowing gas (see Fig. \ref{fig:civspec}). 
        
                \begin{table}[b]
                \centering
                \caption{Inclination (in units of deg) and dynamical mass (in units of \msun) of \1015\ and the \cii\ emitters detected in the ALMA map. For CompB, which is only marginally resolved on one axis in our observation, we assume as size of the \cii\ emitting region $1.5''\times0.18''$, where $0.18''$ is the minor axis of the ALMA beam. The black hole mass of the QSO (see Sect. \ref{sect:mdyn}) is also listed.}
                \makebox[1\columnwidth]{
                        \begin{tabular}{lcccc}
                                \toprule
                                Source       & J1015$+$0020 & CompA  & CompB & CompC \\
                                \midrule
                                $i$ [deg]        & $54\pm12$ &  $55\pm7$ & $-$ & $72\pm3$  \\
                                Log$(M_{\rm dyn}/\rm M_\odot)$  & 10.6$\pm$0.3&  9.4$\pm$0.3 & $>10.7$ & 11.2$\pm$0.2 \\
                                Log$(M_{\rm BH}/\rm M_\odot)$ & 9.8$\pm$0.2& $-$ & $-$ & $-$ \\
                                \bottomrule

                        \end{tabular}
                }
                
                \label{tab:masses}
        \end{table}
        
        \noindent We find that the CIV profile is best fitted by the combination of two Gaussians with dominant contribution from the BLR component. The best fit values are FWHM$_{\rm CIV}^{\rm BLR}=6330\pm270$ \kms and $\lambda L_{\lambda1350 \AA}=(5.0\pm0.8)\times10^{46}$ \ergs. 
        These quantities are used to derive \mbh\ according to the single epoch relation from \cite{Vestergaard06} and considering the empirical correction proposed by \cite{Coatman17} for high-luminosity QSOs to take into account the blueshift $\Delta v_{\rm CIV} = 1050\pm310$ \kms\ affecting the CIV line profile as follows:
        
        \begin{eqnarray}\label{mbh-civ}
        \log{\left(
                \frac{M_{\rm BH}}{{\rm M}_\odot}
                \right)} = 6.71 + 0.53 \log{\left[
                \frac{\lambda L_{\lambda 1350 \AA}}{10^{44}\mbox{ erg s}^{-1}}
                \right]}\nonumber \\
         + 2 \log{\left(
                \frac{{\rm FWHM}_{\rm CIV}}{1000\mbox{ km s}^{-1}}
                \right)} - 2 \log{\left[
                \alpha \left(
                \frac{\Delta v_{\rm CIV}}{1000\mbox{ km s}^{-1}}
                \right) + \beta
                \right]}
        ,\end{eqnarray}
        where $\alpha \sim 0.4$ and $\beta \sim 0.6$. The resulting black hole mass of \1015\ is $5.7^{+3.4}_{-2.1}\times10^9$ \msun, where the uncertainties are dominated by the intrinsic 0.2 dex scatter in the $\Delta v_{\rm CIV}$-corrected relation from \cite{Coatman17}.
        This leads to an Eddington ratio of $\lambda_{\rm Edd} = 0.23^{+0.14}_{-0.09}$.

        \begin{figure}[t]
                \centering
                \caption{Black hole mass as a function of dynamical mass of \1015, compared with a sample of high-z, luminous, and hyper-luminous QSOs from literature. Specifically, the magenta star refers to our target, while diamonds refer to $z\sim4.8-7.1$ QSOs observed in \cii\ with ALMA (see details in text). \mdyn\ are computed according to Eq. \ref{eq:mdyn}, while \mbh\ are single epoch estimates. The best fit $M_{\rm BH} - M_{\rm dyn}$ relation from Jiang et al. (2011) is also indicated by the dashed line with the relative 0.42 dex intrinsic scatter (shaded region).}
                
                \includegraphics[width=0.48\textwidth]{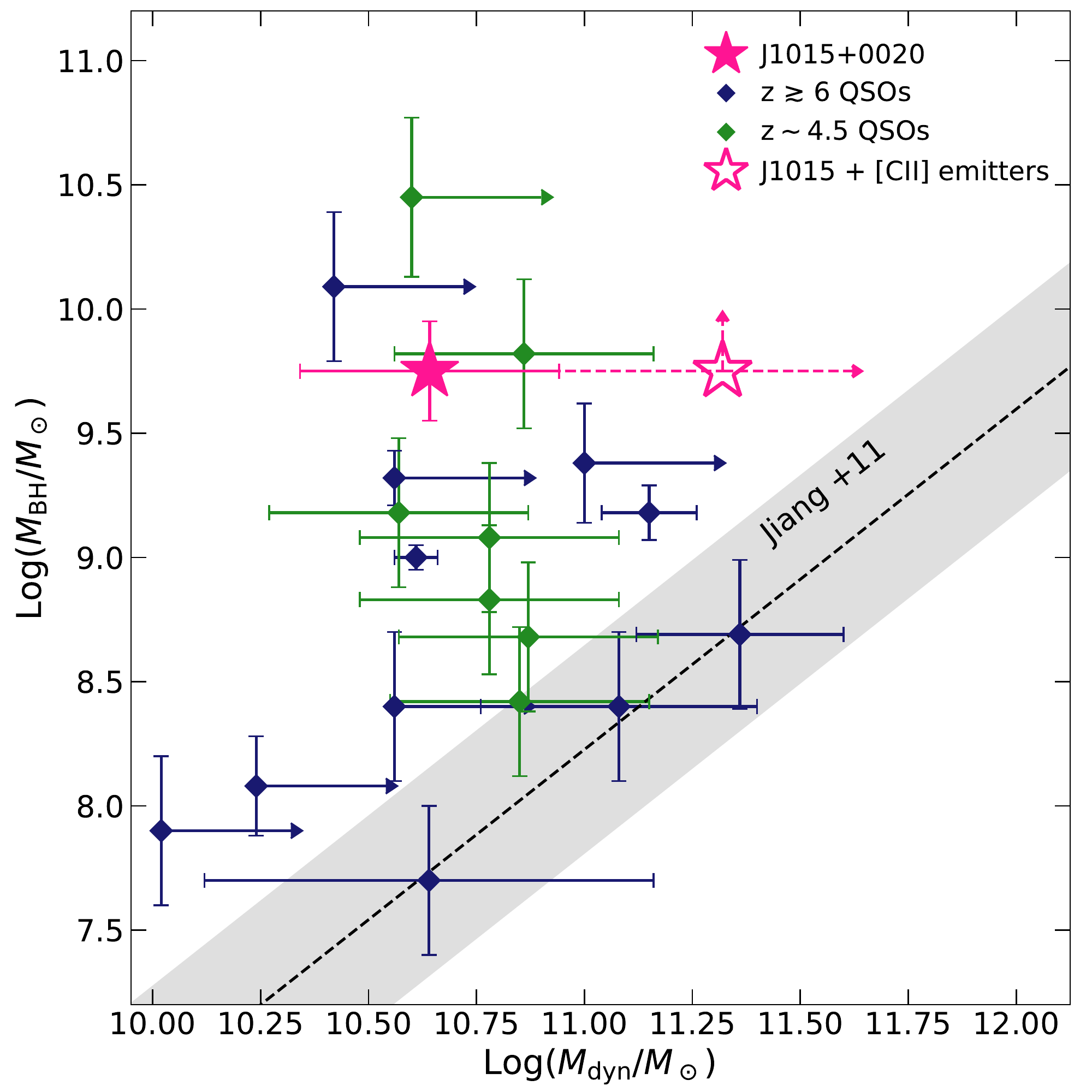}
                \label{fig:mbh-mdyn}
        \end{figure}

        The measurements of the FWHM and size of the \cii\ emission can be used to estimate the dynamical masses of both the QSO host galaxy and CompA, CompB, and CompC. Under the assumption that the ISM is mainly distributed in an inclined, rotating disk, the dynamical mass can be expressed as
        \begin{equation}
                M_{\rm dyn}/M_\odot =9.8\times10^8\left(\frac{D_{\rm [CII]}}{\rm kpc}\right)\left(\frac{FWHM_{\rm [CII]}}{\rm 100 km s^{-1}}\frac{1}{sin(i)}\right)^2 
                \label{eq:mdyn}
        ,\end{equation}

        \noindent where $D_{\rm [CII]}$ is the deconvolved major axis of the [CII]-emitting region, computed as 1.5 times the deconvolved major axis \citep{Wang13}, and $i$ is the inclination angle between the line of sight and the polar axis of the disk. The FWHM of the \cii\ line is related to the circular velocity in the disk by the relation $v_{\rm circ} = 0.75\times \mathrm{FWHM}_{\rm [CII]}/sin(i)$ \citep{Wang13}. In case of a resolved source the inclination of the disk can be estimated from the ratio of semi-minor to semi-major axes as $i=arcos(a_{\rm min}/a_{\rm maj})$. The resulting inclination values for \1015\ and the \cii\ emitters resolved by the ALMA beam are listed in Table \ref{tab:masses}. We also report the statistical uncertainty associated with $i$ derived from the 2D Gaussian fit of the ALMA data. Nonetheless, we point out that inclination estimates of marginally resolved sources (such as \1015\ and CompA), detected at moderate signal-to-noise ratio, 
        can be altered by non-circular beam shapes. In our ALMA observation we might also be losing the more extended, low surface brightness \cii\ emission and, thus, underestimating $D_{\rm [CII]}$. Moreover, the deconvolved size of the source and the semi-axes ratio are estimated from a 2D-Gaussian profile that might not well reproduce the surface-brightness distribution; further discussion on these issues can be found in \cite{Trakhtenbrot17} and references therein.
        Deeper, higher resolution observations are therefore needed to reach a firm conclusion concerning the inclination of the detected sources and, therefore, their dynamical masses.
        Keeping in mind all these limitations, we use Eq. \ref{eq:mdyn} to estimate the dynamical mass of \1015, i.e. Log$(M_{\rm dyn}/\rm M_\odot)=10.6\pm0.3$. We note that assuming the galaxy to be supported by velocity dispersion would lead to a smaller $M_{\rm dyn}$ by a factor of 3.

        \begin{figure*}[t]
        \centering
        \caption{\textit{Panel (a):} black hole accretion rate as a function of the SFR for \1015\ (magenta star) compared to the sample of $z\gtrsim4.5$ QSOs described in Sect. \ref{sect:over} and five WISSH QSOs from \cite{Bischetti17,Vietri18,Duras17}. Orange(red) stars refer to the \cii(continuum) emitters detected in the ALMA map. The locus of points with unitary mass loading factor is also indicated by the dashed line. \textit{Panel (b)}: present BH vs. host galaxy growth timescales, derived assuming constant $\dot{M}_{\rm acc}$ and SFR equal to the observed values. The dotted, dashed, and solid lines indicate a 10:1, 1:1, and 0.1:1 growth timescale, respectively. \textit{Panel (c)}: exponential BH vs. host galaxy growth timescales, derived assuming constant $\lambda_{\rm Edd}$ and SFR/$M_{\rm dyn}$. Lines as in $\textit{(b)}$. }
        \includegraphics[width=0.48\textwidth]{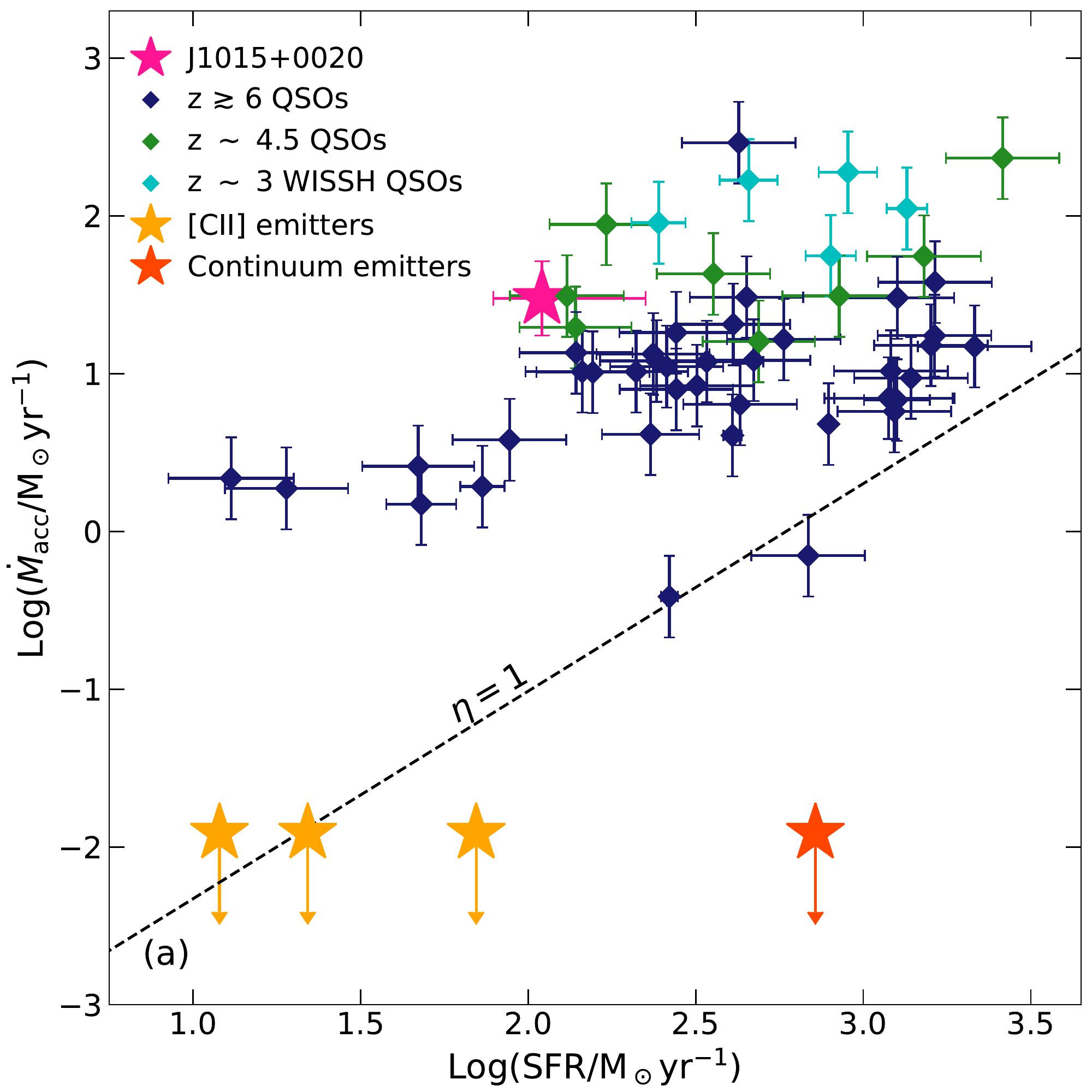}\\
        \includegraphics[width=0.48\textwidth]{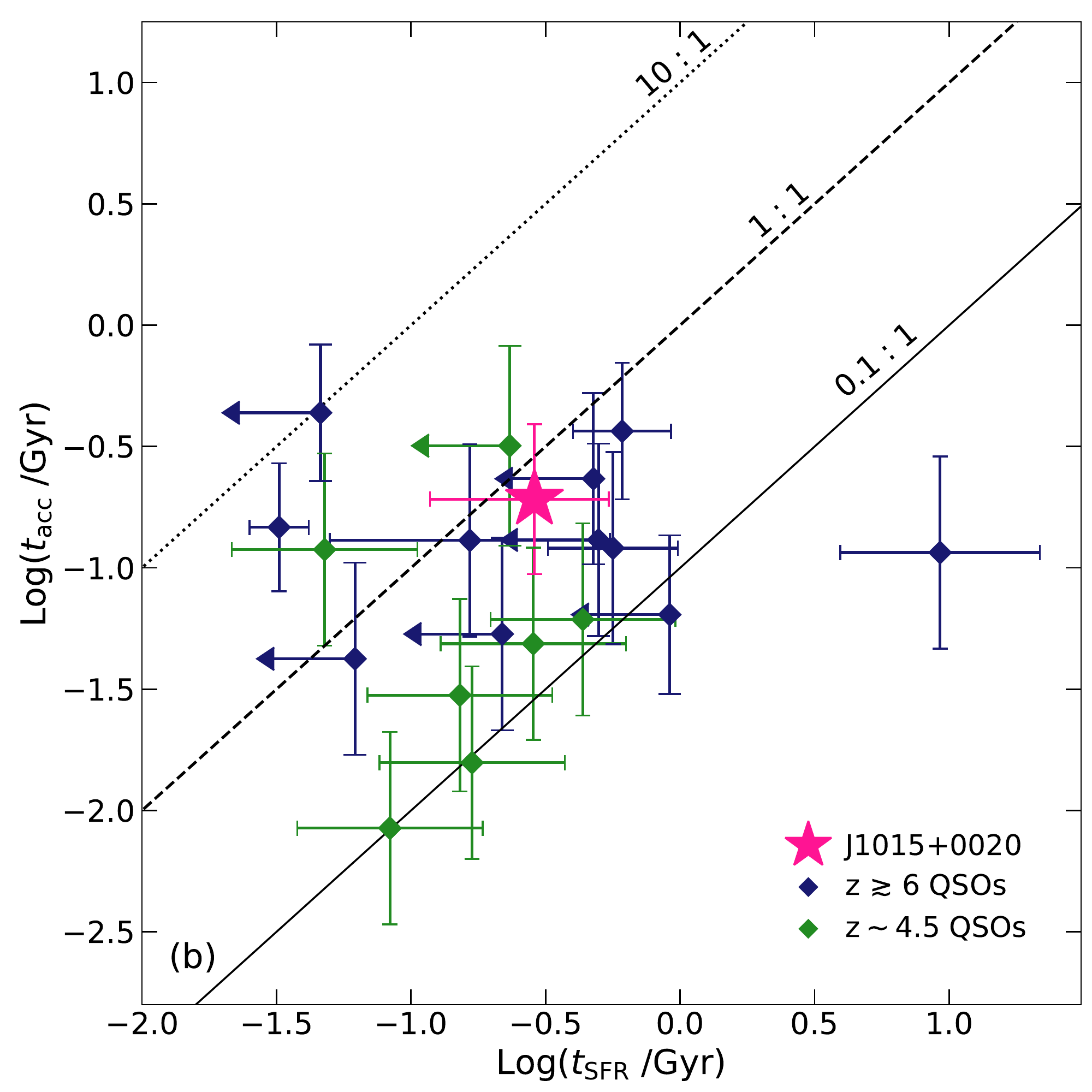}
        \includegraphics[width=0.48\textwidth]{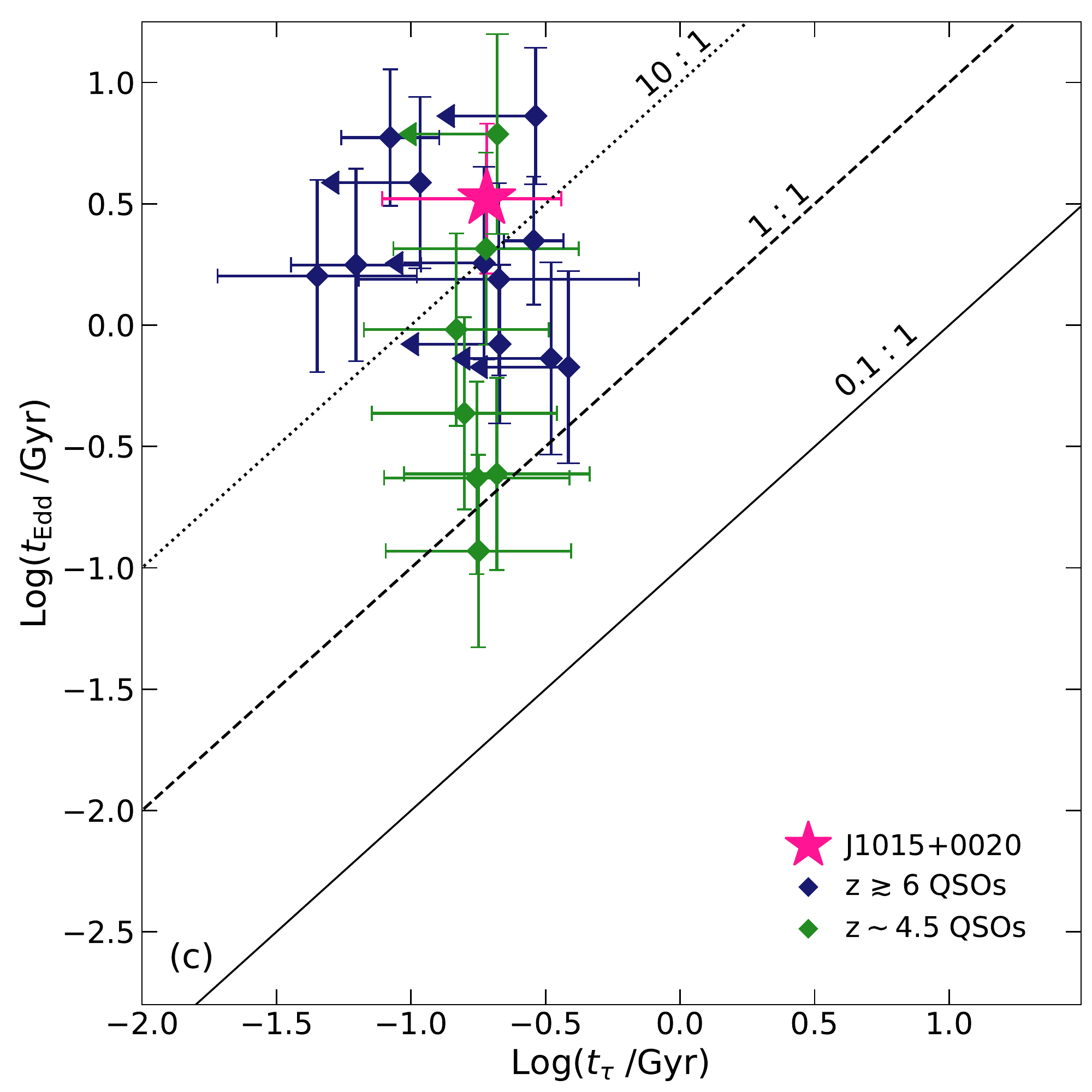}
        \label{fig:mdot-sfr}
\end{figure*}

        \1015 is characterised by an extreme ratio of about $1:7$ of \mbh\ with respect to \mdyn, as shown in Fig. \ref{fig:mbh-mdyn}. Our QSO is compared to the high-z QSOs with single epoch \mbh\ estimates from the sample introduced in Sect. \ref{sect:over}. We also plot the relative uncertainties. Whether these are not available, we consider a 0.3 dex error on the (MgII-based) \mbh\ estimates, while we propagate the inclination error to the uncertainty on \mdyn. For unresolved sources, we plot the \mdyn$sin^2(i)$ value as lower limit on the true dynamical mass. 
        The \mbh-\mdyn\ relation from \cite{Jiang11}, derived from local galaxies in a wide range of \mdyn$\sim10^9-10^{12}$ \msun, is also shown for comparison. According to the local relation, the \mbh-\mdyn\ ratio at the observed black hole mass should be about $1:600$, translating into a host galaxy dynamical mass of $\sim4\times10^{12}$ \msun, suggesting that we are observing the cradle of what would be a giant galaxy at $z = 0$.
        
        As we also resolve CompA and CompC in our ALMA observation, we are able to estimate their dynamical masses Log$(M_{\rm dyn}^{\rm A}/\rm M_\odot)=9.4\pm0.3$ and Log$(M_{\rm dyn}^{\rm C}/\rm M_\odot)=11.2\pm0.2$ (see Table \ref{tab:masses}). CompB is instead only marginally resolved on one axis in our observation and an assumption on $D_{\rm [CII]}$ is therefore necessary. We use $D_{\rm [CII]} = 1.5\times0.18''$ (where 0.18$''$ is the minor axis of the ALMA beam)  and Eq. \ref{eq:mdyn} to derive a lower limit on its dynamical mass of $M_{\rm dyn}^{\rm B}sin^2(i)=5.4\times10^{10}$ \msun.
    Given their small distance from \1015\, all these companions are likely going to merge and eventually build up the mass of the QSO host galaxy. By combining the dynamical masses of the QSO and \cii\ emitters we obtain a large value of $M_{\rm dyn}^{\rm Tot}\sim2.2\times10^{11}$ \msun\ already at $z=4.4$, as shown in Fig. \ref{fig:mbh-mdyn}.

        \subsection{SMBH and host galaxy growth}\label{sect:growth}

        In the previous sections we have discussed the presence of multiple companions at very close distance ($\lesssim23$ kpc) from \1015, which likely contributes to the final mass of the QSO host galaxy. 
        By comparing the mass accretion rate of the SMBH with the SFR, we can in principle understand how high-z QSOs have grown to reach their location in Fig. \ref{fig:mbh-mdyn}, under the assumption that we observe them when most of the black hole and galaxy mass is being assembled.
        
       The value $\dot{M}_{\rm acc}$ can be derived as $\dot{M}_{\rm acc} = L_{\rm Bol}/\epsilon c^2$, once assumed a standard accretion efficiency $\epsilon=0.1$. The resulting value $\dot{M}_{\rm acc}=30\pm7$ \msunyr\ for \1015\ is shown in Fig. \ref{fig:mdot-sfr}a as a function of the SFR, compared to the 42 high-z QSOs sample introduced in Sect. \ref{sect:mdyn}. We also include five WISSH QSOs with available \mbh\ and \lbol\ estimates presented in \cite{Bischetti17,Duras17,Vietri18}.
       If unavailable in literature, we compute \lbol\ by using the bolometric correction from $\lambda L_{1450\AA}$ of \cite{Runnoe12}, assuming as uncertainty the 0.1 dex scatter found for this correction. We also compute the SFR from the \lfir(8$-1000$ \mum) according to the relation reported in \cite{Kennicutt12}, with an associated scatter of 0.2 dex, and dividing the SFR by a factor of two in case of QSOs with \lbol$>10^{47}$ \ergs, as suggested by \cite{Duras17}.
        
        From the relation of \cite{Fiore17} between mass outflow rate and AGN bolometric luminosity Log($\dot{M}_{\rm out}/\rm M_\odot yr^{-1})=0.76\times$Log($L_{\rm Bol}$/erg s$^{-1}$)$-$32, we can define the locus of points in Fig. \ref{fig:mdot-sfr}a with unitary mass loading factor $\eta=\dot{M}_{\rm out}$/SFR, which translates into a $\dot{M}_{\rm acc} = 1.32\times\rm Log(SFR) -3.64$. \1015 and most of the high-z QSOs clearly lie above this line, suggesting that they are potentially able to develop massive molecular outflows affecting the growth of their host galaxies. 
        Future deep ALMA observations of the CO emission in these objects will be able to confirm this prediction.
                 
        From the non-detection of the five QSO companions in the WISE bands, we can compute an upper limit on the \lbol\ of a possible AGN contribution in these sources. 
        Specifically, by using Mrk231 and NGC6420 templates \citep{Polletta07,Fiore08}, we derive a value of Log(\lbol/\ergs) $\lesssim43.84$, corresponding to a $\dot{M}_{\rm acc}\lesssim0.01$ \msunyr. In the companion galaxies, where the AGN effect is absent or very low, it is possible to have star formation activity at a comparable or even higher SFR (orange and red stars in Fig. \ref{fig:mdot-sfr}a, see also Sect. \ref{sect:SFR}) with respect to the QSO host galaxy itself. This suggests that a significant percentage of stellar mass may be assembled in the QSO satellites and then contribute at later times to the QSO host galaxy mass by mergers.

        We compare the SMBH growth and stellar mass assembly timescales in \1015. A basic estimate of the present \mbh\ growth timescale can be derived as the ratio $t_{\rm acc} = M_{\rm BH}/\dot{M}_{\rm acc}$, if a constant mass accretion rate equal to the observed value is assumed. Following the same approach, i.e. by assuming a constant SFR, one can estimate the present stellar mass growth timescale as $t_{\rm SFR} = M_{\rm dyn}/\mathrm{SFR}$. Few combined \cii\ and CO observations of high-z QSOs are available so far \citep{Wang13,Venemans17b}, providing a wide range of molecular gas fractions contributing to the total dynamical mass, from $\sim$10\% to 80\%, if a negligible dark matter content in the inner regions of the galaxy is assumed \citep{Genzel17}. Therefore, the resulting \tsfr\ should be considered as upper limits of the real stellar mass assembly timescales.
        Fig. \ref{fig:mdot-sfr}b shows \tacc\ as a function of \tsfr\ for \1015\ and the same high-z QSOs plotted in Fig. \ref{fig:mbh-mdyn}. The value \tsfr\ ranges from about 30 Myr to 1 Gyr, while the black holes have reached their current mass in $\sim1-300$ Myr. The two growth timescales appear comparable, bearing in mind the large uncertainties affecting these measurements, as most sources lie close to the $1:1$ relation. This suggests that the QSOs of Fig. \ref{fig:mbh-mdyn} are moving in the \mbh-\mdyn\ plane in parallel to the local relation. However, in most of the $z\sim4.5$ QSOs the \tacc/\tsfr\ ratio is $<1$. This is likely due to the extreme accretion rates $\lambda_{\rm Edd}\sim0.8-5$ measured in these sources.

        An alternative approach consists of estimating the exponential growth timescale of a black hole accreting at constant Eddington rate $\lambda_{\rm Edd} = L_{\rm Bol}/L_{\rm Edd}$, where $L_{\rm Edd}$/\ergs\ $= 1.28\times10^{38}$ $M_{\rm BH}/\rm M_\odot$ is the Eddington luminosity. Specifically, following \cite{Volonteri&Rees05}:
        $$t_{\rm Edd}= \tau_{\rm acc} \frac{\epsilon}{1-\epsilon}\lambda_{\rm Edd}\times \mathrm{ln}(M_{\rm BH}/M_{\rm BH}^0)$$
        where $\tau_{\rm acc}\sim0.45$ Gyr is the characteristic accretion timescale \citep{Shapiro05} and $M_{\rm BH}^0\sim10^3$ \msun\ is the initial mass of the black hole seed (see Fig \ref{fig:mdot-sfr}c). We thus calculate the exponential stellar mass growth timescale of the host galaxy by assuming constant $\tau = $SFR/\mdyn\ ratio, equal to the observed value as follows:
        $$ t_{\tau} = \frac{1}{\tau}\left[\mathrm{ln}(M_{\rm dyn}) - \mathrm{ln}(M_{\rm dyn}^0)\right],$$
        where $M_{\rm dyn}^0$ is the initial dynamical mass. Specifically, we compute $M_{\rm dyn}^0 = 9.5\times10^8$ \msun\ from the minimum mass of a star-forming dark matter halo \citep{Finlator11} by rescaling for the cosmological baryon fraction $\Omega_M$, since for the majority of the sources (including \1015) we are not able to distinguish between gas and stellar mass.
	    We find that \tedd\ is on average a factor of 10 larger than \tacc\ (see Fig. \ref{fig:mdot-sfr}c), ranging from 300 Myr to 6 Gyr. This translates into a typical ratio of exponential black hole mass growth to exponential stellar mass growth of $10:1$, as \ttau\ is comparable to \tsfr. 
        According to this scenario, the host galaxies of high-z QSOs have grown in a shorter timescale than their central SMBHs. A shorter BH growth timescale can be obtained by either assuming a more massive seed (e.g. $10^5$ \msun) or a super-Eddington accretion regime onto stellar mass seeds \citep[$\sim$100 \msun; e.g.][]{Volonteri10,Valiante17}.
        
        According to this scenario, the host galaxies of high-z QSOs are growing faster than their central SMBHs. 
        Finally, we note that adopting a $\tau = 2.4$ Gyr$^{-1}$, typical of $z=6-7$ galaxies \citep{Stark09,Gonzalez10}, would shift most QSOs towards larger stellar \ttau, i.e. closer to the $1:1$ relation. This indicates that the bulk of the high-z QSOs sample considered here are undergoing a peculiar phase of intense SF activity.

\section{Conclusions}

In this work, we report on the analysis of the ALMA high-resolution $0.18''\times0.21''$ observation of the 840 \mum\ continuum and [CII] $\lambda157.74$ \mum\ line emission of the WISSH quasar \1015\ and its surrounding (8.5 arcsec radius) field. This data allows us to characterise the host galaxy and environment properties of this hyper-luminous QSO at $z\sim4.4$.
Our main findings can be summarised as follows:
\begin{itemize}
        \item The ALMA observation reveals an exceptional overdensity of [CII]-emitting companions with a very small ($<150$ \kms) velocity shift with respect to the QSO redshift. Specifically, we report the discovery of the closest companion observed so far in submillimeter observations of high-z QSOs. This companion is only 2.2 kpc distance and merging with the QSO, as indicated by the increased velocity dispersion in the host of \1015\ offset from the AGN location. The other two [CII] emitters are located at 8 and 17 kpc.
        \item We also detected two continuum emitters (Cont1 and Cont2 in Table \ref{tab:emission-prop}) within an angular separation of less than 3.5 arcsec, which are characterised by a 840 \mum\ continuum flux density comparable to that of the hyper-luminous QSO. These sources do not show line emission in the ALMA band but are likely physically associated with \1015. In fact, they exceed by a factor of 100 the number of expected sources according to the 850 \mum\ Log(N)-Log(S) (see Sect. \ref{sect:over}). Comparing the five companions detected around \1015\ with typical number density of galaxies from deep submillimeter surveys, clearly indicates the presence of a significant overdensity of star-forming galaxies around this $z\sim4.4$ QSO. We are observing the early phase of the formation of a giant, massive galaxy, assembled by merging of the ALMA-detected companions with a $\sim10^{10}$ \msun\ SMBH at its centre.

        \item We are able to accurately build up the SED of the emission from the QSO, by quantifying and removing the contribution to the 250$-$840 \mum\ fluxes from Cont1 and Cont2 (see Sect. \ref{sect:SFR}). The QSO host galaxy has a SFR of about 100 \msunyr, while the bulk of the SF activity takes place in Cont1 and Cont2, for which we derive a SFR$\sim$700 \msunyr. The \cii\ emitters contribute to additional 100 \msunyr. A significant percentage of the stellar mass assembly at earlier epochs may have therefore taken place in the companion galaxies, more than in the QSO host galaxy itself \citep[e.g.][]{Angles-Alcazar17a}. 
        \item For \1015\ we measure a SMBH mass of $\sim6\times10^9$ \msun\, using a single epoch relation based on the CIV emission line profile, which is a reliable tracer of the mass since it does not exhibit a strong blueshifted wing associated with non-virial motions \citep[][see Sect. \ref{fig:mbh-mdyn}]{Coatman17,Vietri18}. From the \cii\ line profile and emitting region, we also compute the dynamical mass of the QSO host galaxy, \mdyn$\sim4\times10^{10}$ \msun. This translates into an extreme \mdyn/\mbh\ ratio of $\sim7,$ which is a factor of 100 smaller than what typically observed in local galaxies. 
        According to the local relation from \cite{Jiang11}, such a SMBH mass should be hosted in giant galaxy with a stellar mass of $\sim1.3\times10^{12}$ \msun. Remarkably, the total stellar mass of QSO plus [CII]-emitting companions already exceeds $10^{11}$ \msun\ at $z\sim4.4$
        \item \1015, as most of the 47 QSOs at $z\gtrsim4.5$ from the sample described in Sect. \ref{sect:mdyn}, is potentially able to drive a massive molecular outflow affecting the SFR in the host galaxy, according to the relation from \cite{Fiore17} between $\dot{M}_{\rm out}$ and \lbol\ (see Sect. \ref{sect:growth}).
        Dedicated ALMA observations of the CO emission in these objects are needed to further investigate this prediction.
        Fig \ref{fig:mdot-sfr}b and \ref{fig:mdot-sfr}c compare the SMBH versus galaxy growth timescales of high-z QSOs by assuming a constant ($\dot{M}_{\rm acc}$, SFR) and ($\lambda_{\rm Edd}$, $\tau$), respectively. We find that the present growth rate of the host galaxy is comparable to that inferred for the SMBH (Fig. \ref{fig:mdot-sfr}b), while Fig. \ref{fig:mdot-sfr}c suggests that the time necessary to reach the observed \mdyn\ is shorter than the time required to the SMBH to accrete the observed mass.

\end{itemize}

\begin{acknowledgements}

        This paper makes use of the following ALMA data: ADS/JAO.ALMA\#2016.1.00718.S. ALMA is a partnership of ESO (representing its member states), NSF (USA), and NINS (Japan), together with NRC (Canada), MOST and ASIAA (Taiwan), and KASI (Republic of Korea), in cooperation with the Republic of Chile. The Joint ALMA Observatory is operated by ESO, AUI/NRAO, and NAOJ. Funding for SDSS-III has been provided by the Alfred P. Sloan Foundation, the Participating Institutions, the National Science Foundation, and the U.S. Department of Energy Office of Science. The SDSS-III web site is http://www.sdss3.org/.
        SDSS-III is managed by the Astrophysical Research Consortium for the Participating Institutions of the SDSS-III Collaboration including the University of Arizona, the Brazilian Participation Group, Brookhaven National Laboratory, Carnegie Mellon University, University of Florida, the French Participation Group, the German Participation Group, Harvard University, the Instituto de Astrofisica de Canarias, the Michigan State/Notre Dame/JINA Participation Group, Johns Hopkins University, Lawrence Berkeley National Laboratory, Max Planck Institute for Astrophysics, Max Planck Institute for Extraterrestrial Physics, New Mexico State University, New York University, Ohio State University, Pennsylvania State University, University of Portsmouth, Princeton University, the Spanish Participation Group, University of Tokyo, University of Utah, Vanderbilt University, University of Virginia, University of Washington, and Yale University.
        The UKIDSS project is defined in Lawrence et al 2007. UKIDSS uses the UKIRT Wide Field Camera (WFCAM; Casali et al 2007) and a photometric system described in Hewett et al 2006. The pipeline processing and science archive are described in Irwin et al (2008) and Hambly et al (2008). We have used data from the 2nd data release, which is described in detail in Warren et al (2007).
        We thank E. Merlin for the assistance with the HST photometry and M. Ginolfi for the useful discussion on the detection of sources in interferometric images. This research project was supported by the DFG Cluster of Excellence ‘Origin and Structure of the Universe’ (www.universe-cluster.de). C.P. acknowledges support from the Science and Technology Foundation (FCT, Portugal) through the Post-doctoral Fellowship SFRH/BPD/90559/2012, PEst-OE/FIS/UI2751/2014, and PTDC/FIS-AST/2194/2012. S.C. acknowledges financial support from the Science and Technology Facilities Council (STFC). LZ acknowledges financial support under ASI/INAF contract I/037/12/0. CF acknowledges support from  the European Union Horizon 2020 research and innovation programme under the Marie Sklodowska-Curie grant agreement No 664931.

\end{acknowledgements}

\bibpunct{(}{)}{;}{a}{}{,} 
\bibliographystyle{aa} 
\bibliography{biblioCII} 

\end{document}